# Shaking and Tumbling: Short- and Long-Timescale Mechanisms for Resurfacing of Near-Earth Asteroid Surfaces from Planetary Tides and Predictions for the 2029 Earth Encounter by (99942) Apophis.

R.-L. Ballouz[1*], H. Agrusa[2], O.S. Barnouin[1], K.J. Walsh[3], Y. Zhang[4], R.P. Binzel[5], V.J. Bray[6], D. N. DellaGiustina[6], E.R. Jawin[7], J.V. DeMartini[8,9], A. Marusiak[6], P. Michel[2,10], N. Murdoch[9], D.C. Richardson[8], E.G. Rivera-Valentín[1], A.S. Rivkin[1], and Y. Tang[6].

[1] Johns Hopkins University Applied Physics Lab, Laurel, MD, USA.
[2] Université Côte d'Azur, Observatoire de la Côte d'Azur, CNRS, Laboratoire Lagrange, Nice, France
[3] Southwest Research Institute, Boulder, CO, USA.
[4] University of Michigan, Ann Arbor, MI, USA
[5] EAPS, MIT, MA, USA.
[6] Lunar and Planetary Lab, University of Arizona, Tucson, AZ, USA.
[7] National Air and Space Museum, Smithsonian Institution, Washington, DC, USA
[8] University of Maryland, College Park, MD, USA.
[9] Institut Supérieur de l'Aéronautique et de l'Espace (ISAE-SUPAERO), Université de Toulouse, Toulouse, France
[10] Department of Systems Innovation, the University of Tokyo, Japan

*ronald.ballouz@jhuapl.edu

**Abstract**

Spectral characterization of near-Earth asteroids (NEAs) has revealed a continuum of space-weathered states for the surfaces of S-complex NEAs, with Q-class NEAs, an S-complex subclass, most closely matching the un-weathered surfaces of ordinary chondrite meteorites.

Dynamical calculations of the orbital evolution of S-complex NEAs revealed that Q-class NEAs tend to have close encounters with terrestrial planets, suggesting that planetary tides may play a role in refreshing NEA surfaces. However, the exact physical mechanism(s) that drive resurfacing through tidal encounters and the encounter distance at which these mechanisms are effective, has remained unclear.

Through the lens of the upcoming (99942) Apophis encounter with Earth in 2029, we investigate the potential for surface mobilization through tidally-driven seismic shaking over short-timescales during encounter and subsequent surface slope evolution over longer-timescales driven by tumbling.

We perform multi-scale numerical modeling and find that the 2029 encounter will induce short-term tidally-driven discrete seismic events that lead to high-frequency (>0.1 Hz) surface accelerations that reach magnitudes similar to Apophis' gravity, and that may be detectable by modern seismometers. It is still unclear if the shaking we model translates to widespread particle mobilization and/or lofting. We also find there will be a significant change in Apophis' tumbling spin state that could lead to longer-term surface refreshing in response to tumbling-induced surface slope changes. We propose that through these mechanisms, space-weathered S-class asteroid surfaces may become refreshed through the exposure of unweathered underlying material. These results will be tested by the future exploration of Apophis by NASA'S OSIRIS-APEX.

## I. Introduction

Spacecraft exploration of near-Earth asteroids (NEAs) has shown that these objects are covered by a weakly cohesive regolith layer dominated by boulders [Fujiwara et al. 2006, Sugita et al. 2019, Lauretta & DellaGiustina et al. 2019, Arakawa et al. 2020, Walsh & Ballouz et al. 2022]. NEAs interpreted as less space-weathered are preferentially found among objects that can make close encounters with the inner planets, suggesting a connection between those close encounters and a resetting of space weathering on NEA regolith [Binzel et al. 2010, Nesvorny et al. 2010]. A sufficiently close approach of an NEA with a terrestrial planet can lead to a tidal encounter, where tidal stresses from the planet can cause physical changes to the asteroid, including mobilization of its regolith layer [Yu et al. 2014, Kim et al. 2023].

Tidal encounters may provide an explanation for the relationship between stony S- and Q-class asteroids and ordinary chondrite meteorites: Q-class, which have visible-to-near infrared reflectance spectra similar to ordinary chondrites, could be S-complex asteroids that have experienced a recent resurfacing event [e.g., Binzel et al. 2010].

While other physical mechanisms have been shown to be plausible sources for NEA surface refreshing (such as spin-up due to radiative torques, thermal fatigue, and impact-induced seismic shaking [Rivkin et al. 2011, Granvik et al. 2016, Graves et al. 2018, 2019, Binzel et al. 2019, DeMeo et al. 2023]), dynamical calculations indicate that Q-class NEAs are more likely to pass within 16 Earth radii compared to S-class NEAs [Binzel et al. 2010]. However, direct numerical simulations of the tidal encounters of NEAs with terrestrial planets have shown that only close encounters (<3 planetary radii) at moderate encounter speeds lead to global-scale NEA disruption and/or mass shedding [Richardson et al. 1998, Nesvorny et al. 2010, Zhang & Michel 2020]. Therefore, it is unclear if sufficiently close encounters have occurred to resurface the Q-class asteroids to match the spectral observations. This has presented a "resurfacing paradox": a physical mechanism for a tidal-based refreshing of NEAs that is compatible with observations, in particular the Earth-encounter distance at which tidal torques may be effective in the resurfacing process, has remained elusive.

The true maximum encounter distance, $r^*$, at which resurfacing may occur is also debated. Binzel et al. (2010) suggested $r^* \sim 16$ Earth radii based on agreement between dynamical calculations of asteroid encounters with Earth and the statistics of taxonomies within the S-complex compared to expectations based on space weathering and Earth-encounter timescales. Nesvorný et al. (2010) similarly suggest that distant encounters (>3 Earth radii) can resurface S-complex asteroids, though their calculations suggested $r^* \sim 5$ Earth radii based on an estimated space weathering timescale, $t_{SW} \sim 10^6$ year, with $r^*$ inversely proportional to $t_{SW}$. Nesvorný et al. (2010) also noted that $t_{SW} < 10^5$ year would require a maximum encounter distance > 20 Earth radii, which would be implausibly large. However, recent observations of space-weathering features on returned samples do suggest that the $t_{SW} < 10^5$ yr for ordinary chondrites (Jin & Ishiguro 2022). Nevertheless, it is clear that, based on observational evidence and dynamical analyses across multiple studies, r* is greater than the tidal disruption or mass shedding limit.

Here, we approach this long-standing problem by considering new insights into the surface mobility of small bodies due to time-varying tidal forces, excited spin states, rubble pile seismicity, and spacecraft observations of NEA surfaces. Specifically, DellaGiustina & Ballouz et al. [2024] recently showed that rubble-pile binary asteroids may experience continuous tidally driven discrete seismic events that lead to high-frequency (>0.1 Hz) surface accelerations that may be detectable by modern seismometers. The tidal encounters of rubble-pile asteroids with terrestrial planets may lead to a similar type of "quaking" that is sufficiently strong to cause short-term surface refreshing.

Furthermore, recent work on granular flow on Phobos has shown that tidal forces from Mars can cause a time-variable tilting of slopes that induces a creep motion on the surface [Ballouz et al. 2019a]. Creeping is thought to have contributed to the current regolith geomorphology on Bennu as the spin of the asteroid slowly accelerates and slopes fail locally to re-adjust to generate observed terraces [Barnouin et al. 2022] and the surface that has been split into two distinct geologic units [Jawin et al. 2022]. **Thermal cycling has also been proposed to contribute to surface creep on regolith-covered asteroids [Bovie et al. 2023].** A time-variable surface tilting has also been proposed to occur for single or binary NEAs that experience tumbling or non-principal axis rotation [Brack et al. 2019, Agrusa et al. 2022]. For these tumbling NEAs, surface regolith and boulders are subject to Euler accelerations due to changes in the angular velocity in the asteroid-fixed frame. Therefore, if an NEA exists in or transitions to a tumbling rotation state, then it may experience gradual grain motion on its surface, leading to long-term surface refreshing.

Recently, Kim et al. [2023] studied the upcoming close encounter of the S-class asteroid (99942) Apophis with Earth in 2029, demonstrating that a change in the asteroid's spin state could lead to localized regolith refreshing. The Kim et al. [2023] study focused on the potential regolith mobilization associated with the instantaneous change in the spin state, rather than the long-term effects of the encounter. We use Apophis' **6.0** Earth radii close approach in 2029 as an example of an asteroid tidal encounter, in order to evaluate the efficiency of our proposed mechanism in refreshing asteroid surfaces. The efficiency by which Apophis' surface may be refreshed by tidal effects beyond ~3 Earth radii will also provide an important constraint on space weathering timescales for ordinary chondrites and S-class asteroids [e.g., Vernazza et al 2009].

To expand on surface resurfacing process that likely affect tidally-induced NEA surface refreshing, in section 2, we review the observational and theoretical evidence for tidally induced changes in the rotation state of NEAs and present supporting observational evidence for our proposed mechanism. In section 3, we present new simulation results of an Apophis tidal encounter for varying internal structure and attitude at perigee, which show that tidal stresses can trigger short-term quaking events that are detectable by modern seismometers. In section 4, we show how tumbling of asteroids leads to surface conditions that could lead to long-term surface mobility. In section 5, we present a prescription for calculating the mass flow on asteroid surfaces from periodic surface tilting and apply this prescription to outcomes presented in section 3. Finally, in section 6, we discuss the possibility of short- and long-term surface refreshing through seismic shaking and changes in the tumbling state triggered by Apophis's 2029 encounter. We also describe the future work needed to better understand tidally induced surface refreshing on Apophis and other NEAs in general.

## 2. The influence of tidal encounters on asteroid rotation state and surface refreshing: Theoretical expectations and observational evidence

Apophis' encounter with Earth (**6.0 Earth radii at perigee**) in 2029 will provide a test of the effectiveness of tidal encounters at refreshing the surfaces of NEAs. While prior efforts describe varying, sometimes conflicting, predictions for the scale and magnitude of surface alterations on Apophis due to the encounter ([Binzel et al. 2020, Dotson et al. 2022], and references therein), many studies have agreed that the spin state will change [e.g., Scheeres et al. 2005, DeMartini et al. 2019, Benson et al. 2023]. As asteroids have irregular asymmetric shapes, the extent of tidally induced spin state changes is sensitive to the object's attitude at perigee [DeMartini et al. 2019, Benson et al. 2023]. Due to its present tumbling state, it is difficult to predict, with precision, Apophis' attitude during its 2029 close encounter. Therefore, the extent to which its spin state will change is unknown. However, recent analysis by Benson et al.1 2023 showed that the most likely outcome of the 2029 encounter would be an increase in the asteroid's rotational speed, with spin period components $P_{\phi l}$ and $P\psi$ decreasing from their current values of 30.62 and 265.7 h, respectively, to post-encounter values of ~21 and ~185 h. $P_{\phi l}$ is the average precession period of the minimum-inertia (long) axis about the angular momentum vector [Samarasinha & A'Hearn 1991], and $P\psi$ is the rotation period about the long axis. If the encounter induces stronger tumbling or wobbling, then Apophis may experience continued surface alterations as its excited spin state dampens. Thus, we hypothesize that the total effect of surface refreshing may not be immediate but rather through the gradual mobilization of the surface through the tumbling or wobbling of the asteroid.

If tumbling can indeed refresh NEA surfaces, we would expect there to be a correlation between NEA "surface freshness" and their rotation properties. Put another way, if Q-classs are indeed refreshed S-classes, then our proposed mechanism would suggest that Q- and Sq-classes should preferentially tumble compared to S-classes. Sq-class asteroids, like Apophis [Binzel et al. 2019], are objects that have spectral and, presumably, space-weathering characteristics that fall in between the Q- and S-class end members. In order to evaluate the validity of this hypothesis, we cross-referenced rotational data from the Asteroid Light Curve Database (LCDB) [Warner et al. 2021] with the NEO spectral survey MITHNEOS [Binzel et al. 2019, Marsett et al. 2022]. In total, we found that 578 S-classes, 149 Sq-classes, and 90 Q-classes in MITHNEOS have their rotational properties reported in the LCDB. The sub-set of tumblers in each of those spectral classes is a small fraction of their population, with 31 S-classes, 24 Sq-classes, and 18 Q-class asteroids. However, when compared to their respective sub-populations, Q- and Sq-classes are far more likely to be observed to be in a tumbling state compared to S-classes (Table 1), with 20.0 ± 4.2% of Q-classes and 16.1 ± 3.0% of Sq-classes exhibiting tumbling compared to 5.3 ± 0.9% of S-classes. The uncertainties reflect a binomial sample standard deviation. Therefore, Q-classes are approximately 3.7 times more likely to be found in a tumbling state compare to S-classes. In these statistics, we have included all asteroids that could possibly be in a tumbling state; however, only smaller subsets of these objects are evaluated to be tumbling with a high degree of confidence (entries with T, T+, or T? flags, see [Warner et al. 2021]). For the set of "high-confidence" tumblers, the fraction of S-class tumblers shrinks to 2.2 ± 0.6 %, and Q- and Sq-classes are approximately 5 times more likely to be in a tumbling state compared to S-classes.

We apply a two-proportion z-test to evaluate the null hypothesis that the population of Q-class tumblers is randomly sampled from the population of S-class tumblers. Using our population data for S-complex tumblers, we calculate a test statistic of -4.9928. Applying a two-tailed test, we find a p-value $\ll$ 0.001, and, therefore, reject the null hypothesis. We conclude that there is strong evidence that the population of Q-class tumblers is not randomly sampled from the S-complex tumblers. In fact, we find that either the proportion of S-complex tumblers would need to be doubled or, similarly, the proportion of Q-type tumblers halved in order for the null hypothesis to be viable. Based on these statistics, we find that the current observations of the rotational properties of Sq-complex asteroids supports the plausibility of tumbling as a mechanism for NEA surface refreshing.

| | S-classes | Sq-classes | Q-classes |
|---|---|---|---|
| Number in MITHNEOS and LCDB | 578 | 149 | 90 |
| Number of Tumblers | 31 | 24 | 18 |
| Fraction of Tumblers | 5.3 ± 0.9% | 16.1 ± 3.0% | 20.0 ± 4.2% |
| Fraction of Tumblers (High Confidence) | 2.2 ± 0.6 % | 10.7 ± 2.5% | 12.2 ± 3.4% |

**Table 1.** The prevalence of tumblers in the Sq-complex. We cross-referenced the MITHNEOS spectral survey [Binzel et al. 2019, Marsset et al. 2022] with the LCDB [Warner et al. 2021] to compare the prevalence of tumblers in each sub-population. Q-classes are 3.7−5.5 times more likely to be found in a tumbling state compared to S-classes.

We note that even though there is an enhancement in the fraction of Q- and Sq-class tumblers compared to S-classes in a relative sense, the actual fraction of tumblers is small. The relatively low fraction of Q- and Sq-class tumblers suggest that either other mechanisms likely play a role in refreshing S-class NEA surfaces, or that sufficient time has elapsed for their excited spin states to dampen (see Section 5.2 for further discussion on dampening timescales).

Population-wide analysis of NEA orbital properties with spectral class [DeMeo et al. 2023] suggest that changes in asteroid rotation state through YORP spin-up, thermal fatigue, and collisions could contribute to surface refreshing of S-classes [e.g., Graves et al. 2018, 2019]. Another short-term process that could act to refresh NEA surfaces is the sudden change in the surface slopes due to a change in the spin state [Kim et al. 2023].

Our analysis here indicates that tumbling could also contribute to the surface refreshing process through a long-term process. Analytical studies and numerical simulations have shown that tidally induced spin state changes are sensitive to close-approach distance, rotation phase, the asteroid's mass distribution, and pre-encounter spin [DeMartini et al. 2019, Kim et al. 2020, Hirabayashi et al. 2021, Benson et al. 2023]. Therefore, some NEAs that undergo tidal encounters may not actually undergo a spin-state change. This feature of planetary encounters may explain why some S- and Sq-classes are *not* refreshed by tidal encounters, and why all classes of asteroids in the S-complex can be found to have small minimum orbit intersection distances (see Fig. 2b of [Binzel et al. 2019]).

Better characterization of Q-class spin states would help to understand the relative importance of planetary encounters over these other mechanisms. As we previously noted, the magnitude of the spin state change depends on Apophis' rotation phase at the time of perigee [Richardson et al. 1998, Hirabayashi et al. 2021]. As Apophis is slowly rotating and already in non-principal axis rotation [Pravec et al. 2014, Brozovic et al. 2018], it may be more susceptible to our tumbling-refresh mechanism than other fast-rotating NEAs. As described in Benson et al. (2023), the surface acceleration caused by the angular acceleration due to Apophis' complex spin state is on the order of the tidal acceleration at perigee. In the rest of this paper, we evaluate the extent to which surface refreshing may be induced by distant planetary encounters through tumbling.

## 3. Short-Term Effects: Direct Simulations of the Apophis Encounter

For this work, we complement previous studies of the potential outcome of the Apophis tidal encounter in 2029 by considering the influence of different internal structures for the asteroid. Apophis' 2029 encounter is a convenient test case for considering population changes as a whole. Analysis of the delay-Doppler radar images of Apophis by Brozovic et al. (2018) suggests that Apophis is bi-lobed, perhaps even a contact binary. What is known of Apophis' prolate shape is reminiscent of the S-class asteroid Itokawa, visited by the Hayabusa sample-return mission [Fujiwara et al. 2006]. To model its rotational acceleration, thermophysical analysis of Itokowa's elongated and bilobate shape has shown that it may have a heterogeneous interior [Lowry et al. 2014]. Lowry et al. (2014) and Kanamaru & Sasaki (2019) have suggested that Itokawa may be composed of a denser "head" region separated from its "body" by a compressed "neck" region.

This internal heterogeneity may be needed to explain the 16–21 m offset between Itokawa's center-of-mass (COM) and center-of-figure (COF), if the surface is assumed to have reached a relaxed geopotential state [Lowry et al. 2014, Kanamaru & Sasaki (2019)]. This COM-COF offset corresponds to a head bulk density, $\rho_{\text{head}}$ = 2.75–2.85 g/cm$^3$, which is 40–60% greater than its less dense body with bulk density of $\rho_{\text{body}}$ = 1.75–1.95 g/cm$^3$ [Lowry et al. 2014, Kanamaru & Sasaki (2019)]. From a structural perspective, these possible differences in head and body may be due to the presence of a largely intact monolithic core in the "head" region of Itokawa [Kanamaru & Sasaki (2019)]. We note that alternative models for the internal properties of Itokawa also exist [e.g., Vokroihlicky et al. 2015].

In order to assess the effect of a similar type of interior heterogeneity on Apophis due to its 2029 close encounter, Benson et al. (2023) analyzed the accelerations, at perigee, on an Apophis-sized body made up of two spherical components. Benson et al. (2023) found that the tidal forces could relieve compression on the neck region such that the asteroid could experience some internal shifting due to the rolling of one component (e.g., the body) relative to the other (e.g., the head) caused by the tangential component of the rotational acceleration. Though, this effect depends on whether Apophis' long axis is aligned or mis-aligned with the vector pointing to Earth at perigee.

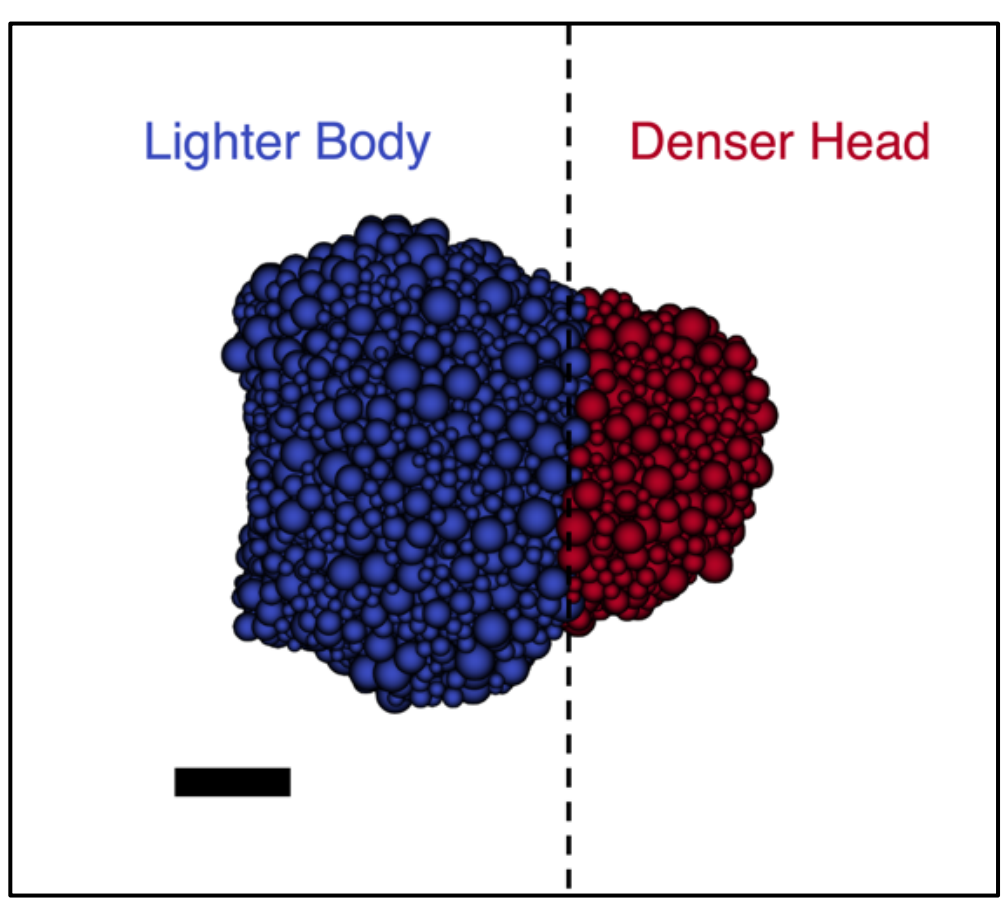

**Figure 1.** Schematic of rubble-pile model of Apophis using the Brozovic et al. (2018) shape model, including the effects of internal heterogeneity by setting head particles (red) to be denser than body particles (blue) by up to 70%. The black bar is a 100-m scale bar.

In order to better understand the possible influence of internal heterogeneity on the outcome of the Apophis 2029 encounter, we model the asteroid's close approach with Earth by using a soft-sphere discrete element code (SSDEM), *pkdgrav*, in a manner similar to previous studies of asteroid tidal encounters with terrestrial planets [Walsh & Richardson 2008, DeMartini et al. 2019, Zhang & Michel 2020] (see Section *A1* for details). We use the Brozovic et al. (2018) radar shape model of Apophis to construct a rubble pile asteroid made up of approximately 4,600 spherical particles. We use a power-law size distribution of randomly packed particles with diameters between 10 and 30 m and a cumulative power-law exponent of –3, based on observations of boulders on the surfaces of km-scale NEAs [DellaGiustina et al. 2019a, Sugita et al. 2019]. As Apophis' mass is not yet well constrained, we have modeled a homogeneous Apophis rubble pile that has a bulk density of 2.2 g/cm$^3$, based on the range in bulk densities of other S-class asteroids visited by spacecraft [Yeomans et al. 2000, Fujiwara et al. 2006, Cheng et al. 2023]. We then constructed a hypothetical heterogeneous Apophis rubble pile that has a denser head and a lighter body (see Fig. 1), with the transition between these two regions occurring at a neck area that was identified through visual inspection of the radar shape model (dashed line in Fig. 1). We consider three cases for an Apophis rubble pile with different internal mass distributions (mass and bulk density is constant across all cases): i) a homogeneous case, ii) a case with $\rho_{\text{head}}$ = 2.8 g/cm$^3$, representing a 37% increase in bulk density between head and body and a COM-COF offset of 9.7 m, and iii) a case with $\rho_{\text{head}}$ = 3.3 g/cm$^3$, representing a 70% increase in bulk density between head and body and a COM-COF offset of 17.8 m. Table A2 summarizes the properties of these three different cases. The latter case represents a likely upper-limit for this configuration, where the head is mostly composed of a single mass concentration, as the bulk densities of ordinary chondrites are at most 3.18, 3.3, and 3.35 g/cm$^3$ for LL, L, and H chondrites, respectively [Flynn et al. 2018]. The pre-encounter rotation state is taken from the radar solution [Brozovic et al. 2018] and the Earth-encounter conditions were taken from the JPL Horizons ephemerides, similar to other works [e.g, DeMartini et al. 2019, Kim et al. 2023, Benson et al. 2023]. Here, we vary the internal heterogeneity in order to explore the extent of possible changes in spin state outcomes, rather than attempt to be prescriptive of the true final spin state of Apophis after its 2029 encounter.

We ran a suite of 54 simulations varying the internal structure of Apophis (3 configurations) and its orientation at perigee (18 evenly spaced configurations), as the outcome of a tidal encounter is sensitive to a body's relative orientation [Scheeres et al. 2005], which is still unknown for Apophis [Durech et al. 2024, T-5y].

### *3.1 Short-Term Effects: Influence of Internal Heterogeneity on Changes in the Rotation State*

For the three different cases of Apophis internal structure, we simulated the 2029 close approach encounter for a duration of ~8 hours (approximately ±4 hours from perigee with Earth) for different initial orientations. Fig. 2 shows the change in the Apophis spin period as a function of $\phi$, defined as the angle between the Apophis COM to Earth COM vector and the asteroid's long axis at Apophis' perigee. The sensitivity of the spin period change to orientation has been discussed previously [DeMartini et al. 2019, Benson et al. 2023] but here we show the nature of this variability as a function of $\phi$. In general, we find that the smallest changes in spin period occur close to $\phi = 0°$ or 90°, and that the spin period increases for $0° > \phi > 90°$, and decreases for $90° > \phi > 180°$, regardless of the asteroid's internal configurations. Overall, there is less variability in the magnitude of the spin-up outcomes (decreasing spin periods) than there is for the spin-down outcomes (increasing spin-periods), consistent with the results of DeMartini et al. (2019).

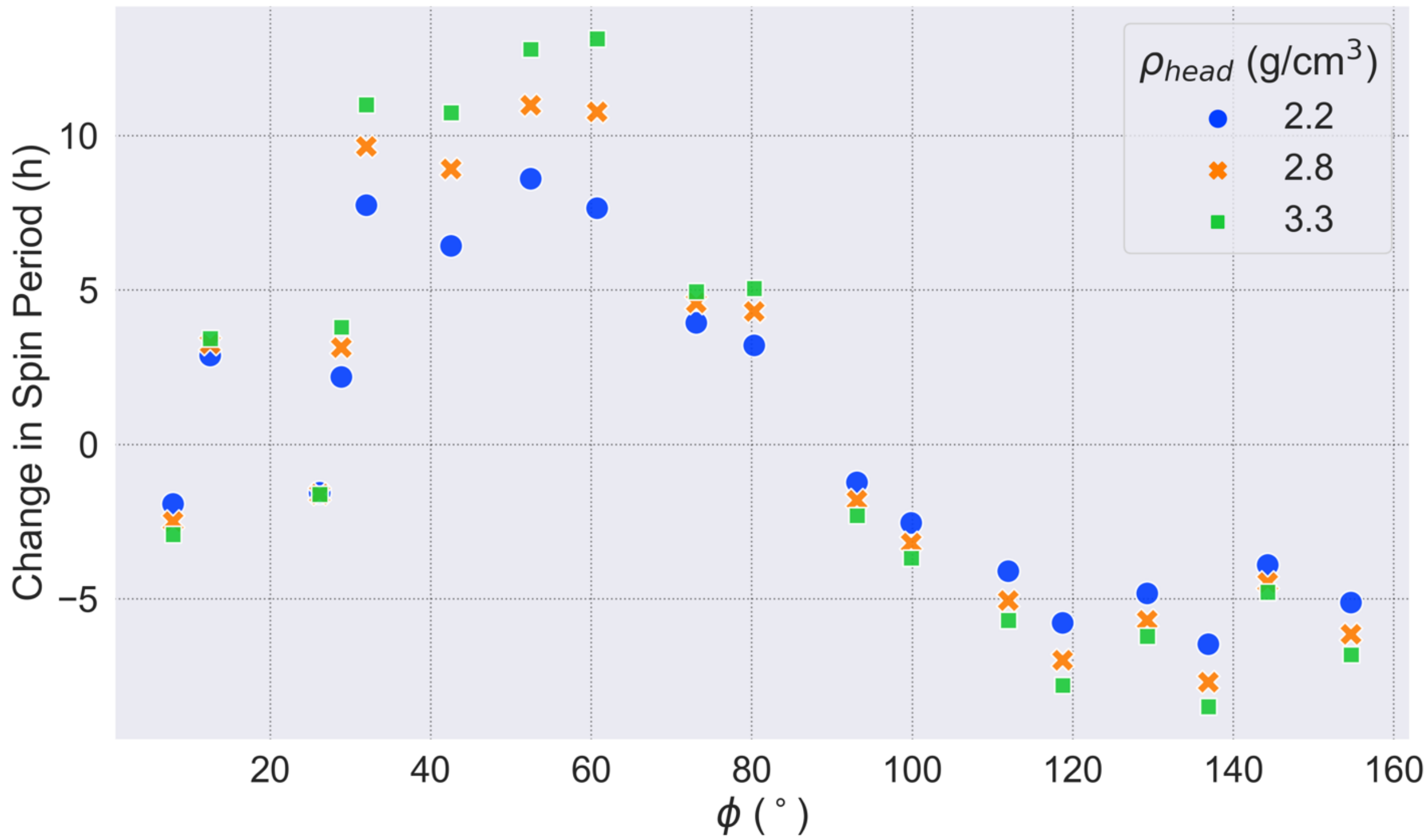

**Figure 2.** The change in Apophis' spin period due to its 2029 Earth tidal encounter is shown as a function of the angle between the Apophis-COM-to-Earth vector and the asteroid's long axis, $\phi$, at perigee and for different asteroid internal structures: homogeneous with $\rho_{\mathrm{head}}$ = 2.2 g/cm$^3$ (blue disks), heterogeneous with $\rho_{\mathrm{head}}$ = 2.8 g/cm$^3$ (orange crosses), and heterogeneous with $\rho_{\mathrm{head}}$ = 3.3 g/cm$^3$ (green squares).

For each configuration of $\phi$, we find that cases with a larger $\rho_{\mathrm{head}}$ (larger COM-COF offset) result in larger changes in the spin period, regardless of whether Apophis ends up spinning up or spinning down. This is likely due to the larger torques applied to the asteroid for cases with a larger COM-COF offset. The largest increase in spin period is found for the $\phi$ = 61° case, where the difference in the spin-period change between the $\rho_{\mathrm{head}}$ = 2.2 g/cm$^3$ and 3.3 g/cm$^3$ cases is +5.5 h. For cases that lead to a spin-up, the largest change was found for the $\phi$ = 137° case, where the difference in spin-period change of the aforementioned cases is –2 h. The smallest change was for $\phi$ = 26° case, where the difference for those same cases is only –3.2 min.

These results illustrate how the characterization of the evolution of Apophis' spin before and after its 2029 encounter with Earth would enable a characterization of the asteroid's interior structure, as described in a continuum-based study by Dinsmore & deWit 2023 and demonstrated for other asteroids [Takahashi et al. 2013]. For the end-member case of a contact-binary Apophis with an L ordinary-chondrite-like $\rho_{\mathrm{head}}$, the change in spin period could influence the exact surface dynamics that could lead to instantaneous resurfacing through a surface-slope change [Kim et al. 2023]. However, it may be difficult to distinguish different internal structures from the change in the spin state alone (e.g., the $\phi$ = 26° case described above). We use the range of outcomes described here to evaluate the long-term effects of tidally induced tumbling on surface refreshing in Section 5.

*3.2 Short-Term Effects: High-frequency Seismic Shaking During Close Encounter*

DellaGiustina & Ballouz et al. (2024) used *pkdgrav* to show that tidal forcing in a rubble-pile asteroid binary system produces seismicity that should be detectable by modern seismometers. In conjunction with periodic elastic deformation, DellaGiustina & Ballouz et al. (2024) observed stochastic events of variable amplitude and length that correspond to anelastic behavior of the interior particles as they shear against each other, leading to interior jostling. *pkdgrav*'s SSDEM functionality allows for the modeling of realistic friction forces that cause tidal dissipation. The combination of normal and tangential restoring and frictional forces enables accurately represents interior shearing forces as an asteroid is deformed by tides, leading to discrete "seismic events" caused by slip-stick phenomena.

Here, we build off the work of DellaGiustina & Ballouz et al. (2024) by evaluating seismicity for a rubble-pile Apophis. We simulated two high-temporal-resolution cases for two different internal structure end-members: a homogeneous case and a heterogenous case with $\rho_{\text{head}}$ = 3.3 g/cm$^3$, with similar initial spin states and orientations. For these cases, we simulated the settling of the rubble piles for 3.6 days (approximately 3 rotation periods) to minimize any small-scale relative motions of the constituent particles. After this simulated settling period, we observe that the motion of surface particles relative to the rubble pile COM is on the order of a few nanometers/s$^2$ (Fig. A1). The tidal encounter simulations are then initialized with the force history of the settled rubble piles preserved, and Apophis is placed on its close-encounter trajectory with Earth. We then simulated ~16 h (approximately ±8 h from perigee) of the tidal encounter. The resulting radial (up-and-down) motion of Apophis surface particles is recorded by computing a 4$^{\text{th}}$-order finite difference of their velocities with respect to the rubble-pile COM with time.

The resulting radial acceleration of a surface particle is show in Fig. 3, parameterized by the acceleration, *a*, normalized by the mean local gravity, *g*, on Apophis for $\rho_{\text{bulk}}$ = 2.2 g/cm$^3$. Fig. 3a shows the outcome for the homogeneous case and Fig. 3b shows the outcome for the heterogeneous case. Each spike in Fig. 3 is a discrete "quaking" event that is induced by the tidal encounter. Surprisingly, we found that Apophis' tidal encounter can excite high-frequency shaking to magnitudes that reach levels of Apophis' self-gravity (cyan-shaded region in Fig. 3a). The elastic response of the rubble-pile due to tides can be observed in the steady rise and subsequent drop of the baseline ground motion (Fig 3). The encounter leads to a raised background shaking level, by an order of magnitude. We highlight three specific quaking events in Fig. 3 that are shown in more detail in Fig. A2, which show that the duration of "strong" shaking events ($|a/g| \gtrsim 0.1$) can be ~1.5 minutes up to 15 minutes. Note that a single $|a/g|$ = 0.1 event is three orders of magnitude larger than the baseline quaking level of settled surface particles on the simulated rubble-pile. We note that the *pkdgrav* timestep is ~ 0.01 seconds, so each of these "shaking" events captures motion at high frequences (~100 Hz). Relatively strong shaking events persist after the close encounter. These are triggered by the internal rearrangement of the rubble-pile structure as it settles into a new post-encounter equilibrium (see Fig. A3).

Similar to the quaking events observed in DellaGiustina & Ballouz et al. (2024), we surmise that the transient events observed in our simulations are due to the buildup of stresses that cause failure in near-surface and sub-surface layers, analogous to localized sources generated by faulting ("quake"') that are examined in conventional seismological studies. To pursue this idea further, we evaluated ground motion globally on the simulated rubble piles.

Fig. 4 shows time-sequence maps of the ground-motion magnitude on Apophis for the case shown in Fig. 3a. Fig. 4 illustrates how quaking on Apophis is initiated at a specific source location and propagates radially outward, followed by "ringing" as the seismic waves are able to traverse across the surface and interior of the asteroid before dissipating. The possible interference of seismic waves in a relatively confined small NEA shows the complex nature of seismic wave propagation on these objects compared to larger planetary interiors that have been characterized with seismometers [Lognonné et al. 2019, Garcia et al. 2022]. The geometrical spreading of the seismic waves is reminiscent of modeled impact-induced seismicity on other NEAs, such as Bennu [Quillen et al. 2019]. Fig. 4e shows that high-amplitude events ($|a/g| > 0.1$) are seen globally on the asteroid. Figs. 4f–h shows how global shaking decays over the next ~2 minutes. The magnitudes of these seismic events are surprisingly large compared to the magnitude of the tidal acceleration at perigee, and suggests that the extent of surface alteration may be greater than had been considered in previous studies [Yu et al. 2014, DeMartini et al. 2019, Kim et al. 2023]. In Section 6, we discuss the potential for mass mobilization on Apophis if seismicity of these magnitudes is indeed induced on the asteroid during its close encounter.

In Fig. 5, we show the power spectral density calculated for the case shown in Fig. 3a (blue curve) compared to the sensitivity of a short period (SP) seismometer developed by Silicon Audio (SiAU) and University of Arizona (green curve) [Marusiak et al. 2020, 2021] and the Insight SIAS SP (yellow curve) and Very Broad Band (VBB, red curve) seismometers [Mimoun et al. 2016, Lognonné et al. 2019]. Fig 5a shows the predicted ground motion on the surface along the asteroid's major axis, and Fig. 5b shows the same for the asteroid's intermediate axis. For the simulation output shown here, the asteroid's intermediate axis is in tension, which produces a stronger signal at lower frequences (~$10^{-2}$ to $10^{-1}$ Hz) compared to the orthogonal axes that are in compression (the major and minor axes). The response of the surface along the minor axes is similar to that shown in Fig. 5a (major axis), as it is also orthogonal to the vector from the asteroid to the Earth at perigee. This fully-body deformation of the asteroid during the close encounter is also illustrated in Fig. A4. Fig. 5 shows that characterized self-noise of the SP seismometers (red and purple curves) is orders of magnitude smaller than the predicted ground motion on the surface of Apophis (blue curve) across a broad range of frequencies, ~$10^{-2}$ Hz up to 0.5 Hz (our simulation output Nyquist frequency), which largely falls between the terrestrial New High Noise Model and New Low Noise Model [Peterson 1993] (signifying that it is within the measurable range of modern seismometers). This high-frequency response is related to the discrete events, similar to those highlights in Fig. 3. In general, there are clear seismic signatures related to the close approach that could be detected by SP or VBB seismometers.

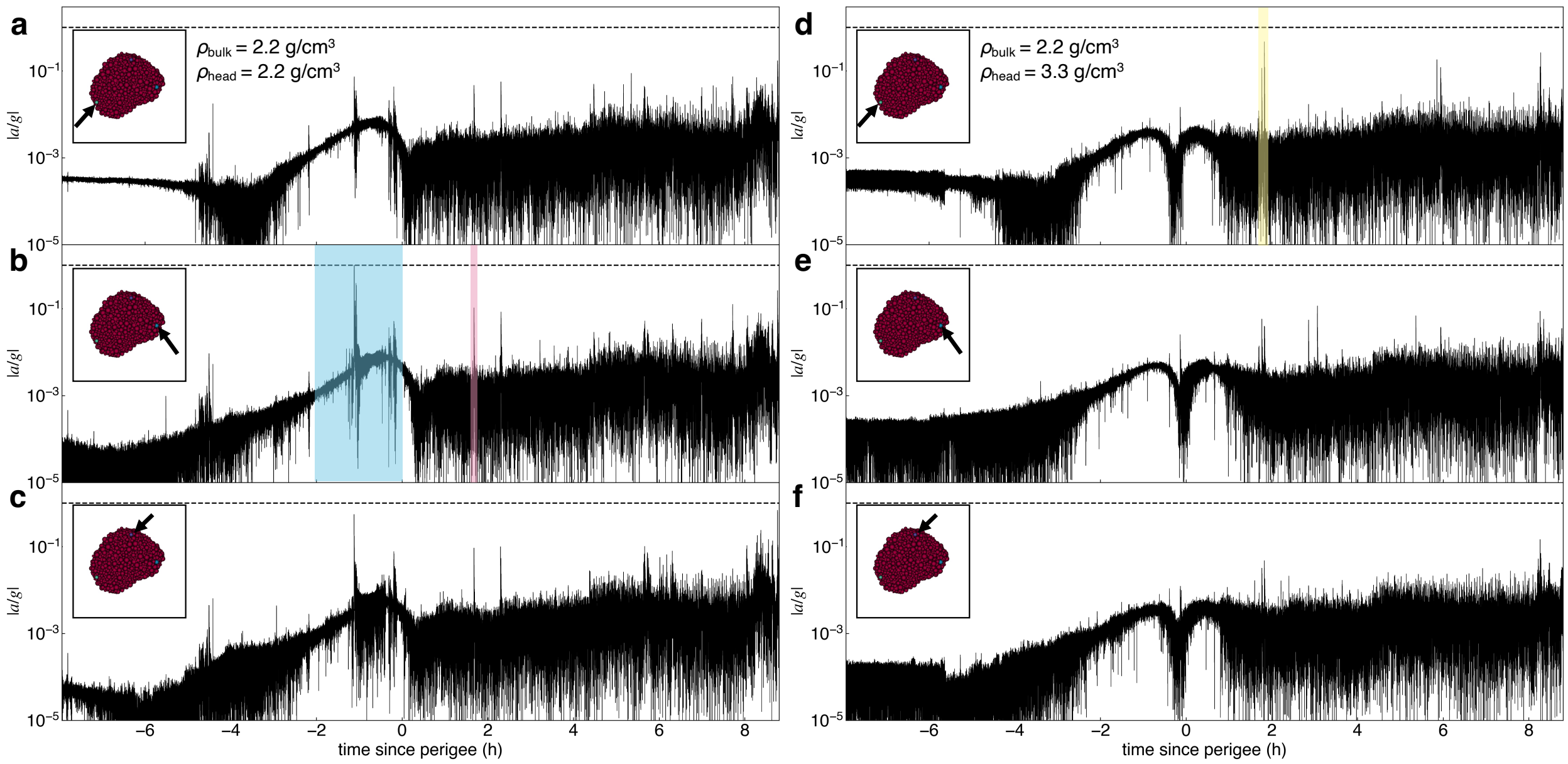

**Figure 3**. Simulated ground motion on the surface of Apophis during its closest approach with Earth on April 13, 2029. Each panel shows the surface radial acceleration, *a*, normalized by Apophis gravity, *g*, for the particle indicated by the black arrow in the inset figure. The brighter-colored particles in the inset align with Apophis' major axes. For each case, the particle exhibiting the largest activity is highlighted in the sub-panel. **a–c**, |a/g| for three surface particles for the case with a homogeneous interior with bulk density 2.2 g/cm$^3$. The cyan-shaded region highlights in **b** shows a discrete shaking event where ground motion reaches a peak of |a/g| ~ 1 (Fig. A2a), and the magenta-shaded region in **b** highlights an event with a peak |a/g| ~ 0.07 (Fig. A2b). **d–f,** same as **a–c**, except the simulated Apophis has a contact-binary-like interior with a denser "head" region that creates a center-of-mass to center-of-figure offset of approximately 17 m. The yellow-shaded region in **d** highlights an event with a peak |a/g| ~ 0.5 (Fig. A2c). The ground motion has a distinct double-peaked feature around perigee, which is different from the homogeneous case. The double peak feature arises from a re-alignment of different body axes with the tidal force vector. As Apophis flies by Earth, the alignment of different body axes with the tidal force flips. Axes that were once in tension become compressed and vice-versa (Fig. A4). This re-alignment is less severe for the homogenous case, which is why the double-peak feature is less pronounced.

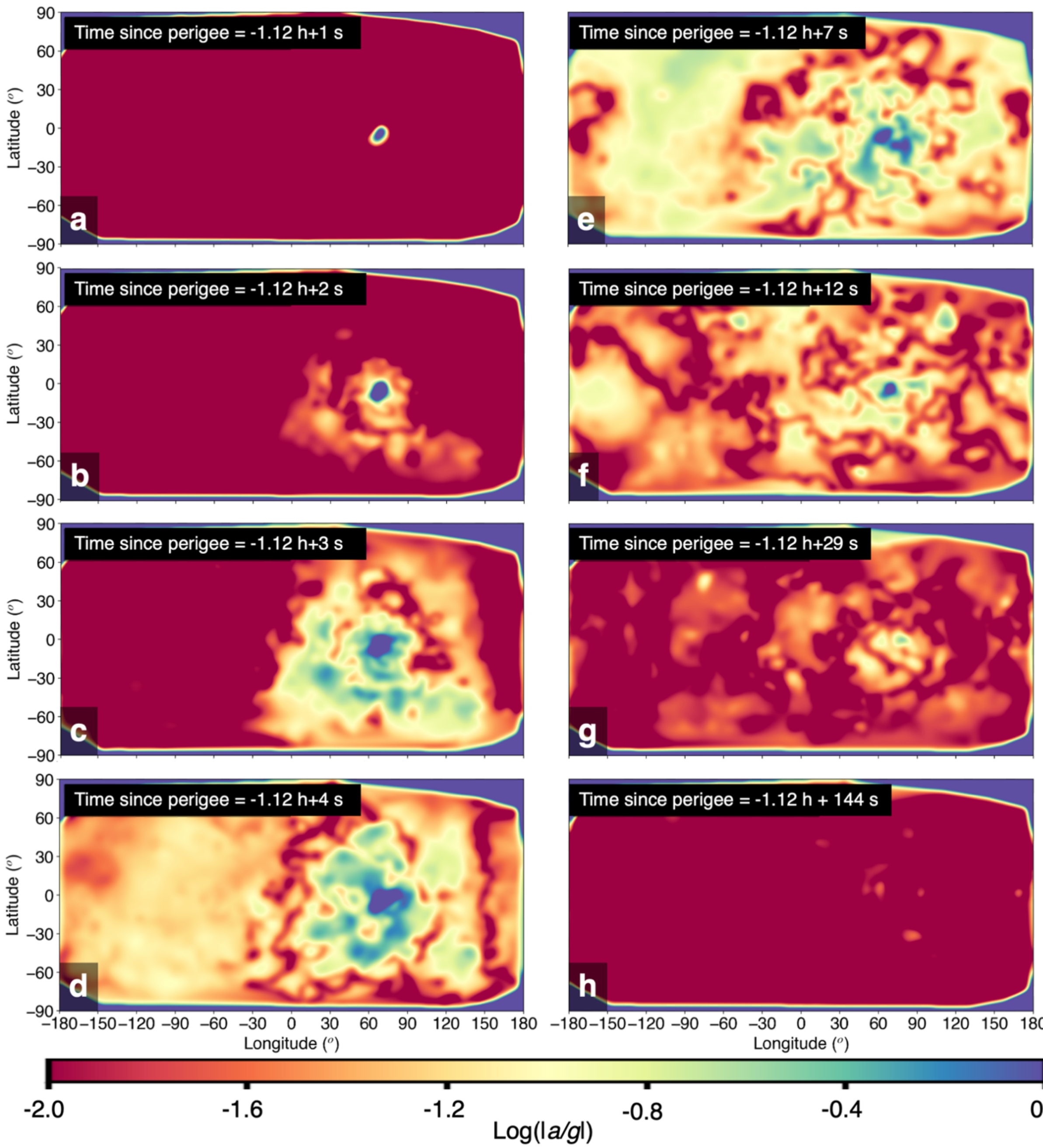

**Figure 4.** A series of snapshots of the surface radial acceleration across the surface of a simulated rubble-pile Apophis during a discrete shaking event triggered by its close encounter with Earth (magenta-shaded region in Fig. 3). The radial accelerations of discrete surface particles are projected to an equatorial map. The bluer color on the edges of each panel do not represent any particles, and are an artifact of the projection of 3D data onto these 2D maps. **a,** the beginning of the event, t = –1.12 h before perigee, where a region near the equator and longitude ~60° W exhibits strong vertical motion that begins propagating radially outward. **b–d**, surface acceleration maps show a seismic surface wave expand radially outward at ~100 m/s. **e,** peak ground motion as Apophis experiences global shaking. Simulations show the possibility for antipodal focusing and ringing as strong shaking ($|a/g| > 0.1|$) continues for 10's of seconds. **f–h**, global shaking decays over the next ~2 min.

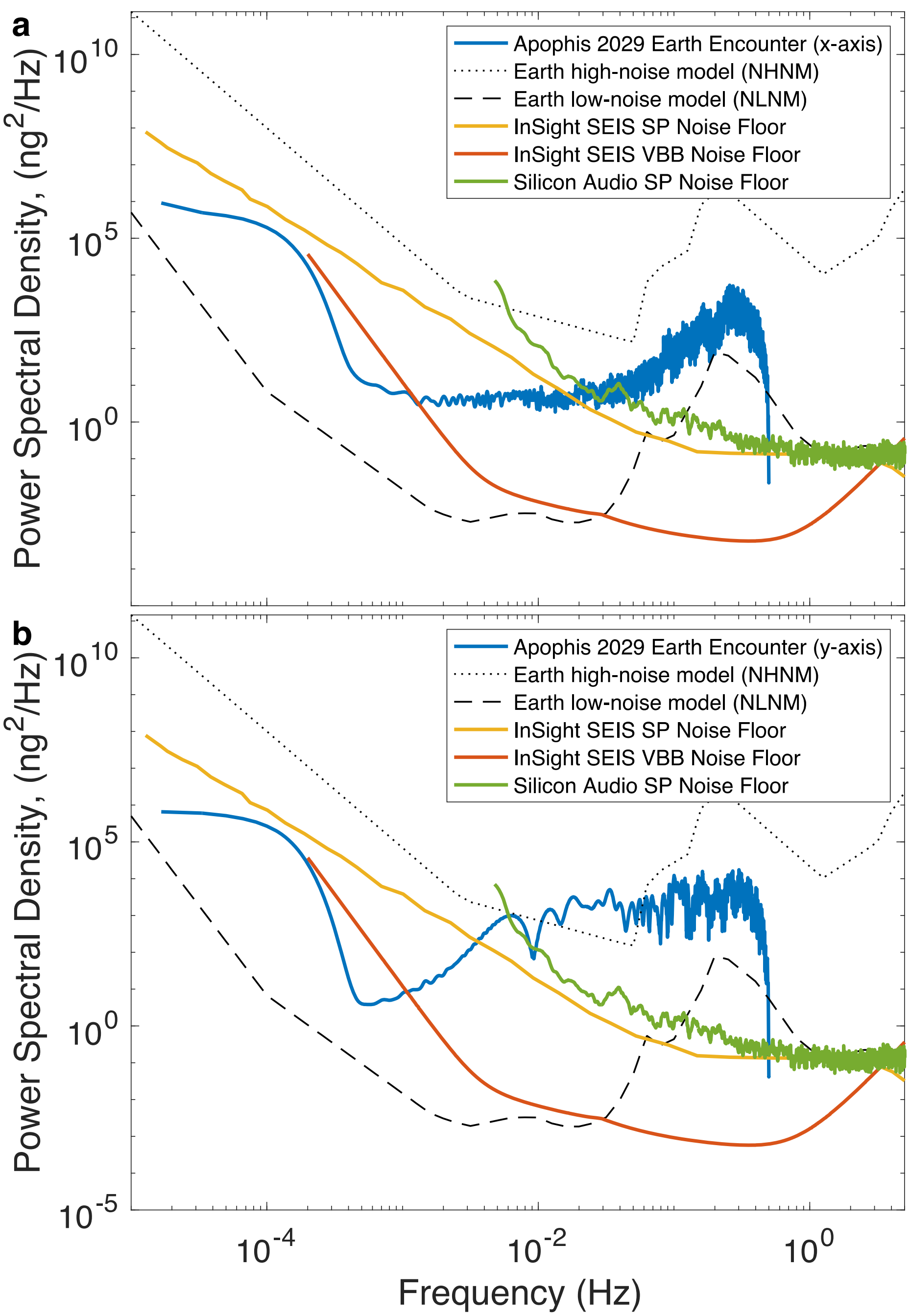

**Figure 5. a,** Self-noise of Silicon Audio SP (green curve) and InSight SEIS SP (yellow curve) and VBB (red curve) seismometers compared to the power spectral density (PSD) computed from the motion of a particle on the surface of Apophis that is aligned with its major axis (x-axis, blue curve). The time series used to compute the ground motion (Fig. 3a) was from 16 h surrounding its close approach with Earth (~**6.0** Earth radii) on April 13, 2029. For reference, the New High Noise Model and New Low Noise Model for Earth [Peterson 1993] are shown (black dashed lines). **b,** same as **a** but the blue curve now showing the predicted motion on the surface on the asteroid's intermediate axis (y-axis) that is aligned with the Apophis-Earth vector at perigee. In both panels, the cutoff at larger frequencies is artificial, representing the simulation output Nyquist frequency. The true response of the asteroid to tidal forces may continue to rise at higher frequencies.

## 4. Long-Term Effects: Tumbling leads to time-varying surface slopes

This study and others have shown that tidal encounters can initiate tumbling or change the tumbling state for asteroids, particularly for slow rotators like Apophis [e.g., Benson et al. 2023]. A tumbling rotation state can lead to time-varying surface slopes on a small body [Brack & McMahon 2019, Agrusa et al. 2022]. Taking the results of the numerical simulations described in Section 3, we calculate the time-varying surface slopes on Apophis by using its radar shape model and an assumed density (2.2 g/cm$^3$). Section A2 details how dynamic surface slopes are calculated for a tumbling asteroid. As shown in Section A2, time-varying surface slopes are a result of a dynamic non-zero Euler acceleration term.

Unlike the work of Brack & McMahon (2019) that considered the mobilization of larger boulders and how their possible lofting could lead to tumbling, we do not include the influence of Coriolis acceleration. Here, we investigate how slope destabilization could possibly lead to a slow creep of material on the surface with displacements that are very small compared to the size of the asteroid [Ballouz et al. 2019a, Jawin et al. 2020]. To compute gravity at each facet, we use the algorithm of Werner & Scheeres (1996) as implemented in Barnouin et al. (2020) with the radar shape model of Apophis [Brozovic et al. 2018]. We focus on the case of a homogeneous Apophis with $\rho_{\mathrm{bulk}}$ = 2.2 g/cm$^3$. To illustrate the influence of tumbling on surface slopes, we consider Case 15 (see Table A3), which led to Apophis tumbling with post-encounter major-axis, intermediate-axis, and minor-axis spin periods of $P_z$, $P_y$, and $P_x$ = 27.6 h, 32.5 h, and 493.4 h, respectively. We choose Case 15 as it presents an interesting possibility where the magnitude of the major- and intermediate-axis spins are most similar compared to other cases with $\rho_{\mathrm{head}}$ = 2.2 g/cm$^3$. Integrating the rigid-body dynamics of an Apophis tumbling with aforementioned spin periods, we solve for the surface slopes on Apophis over a tumbling period.

Fig. 6a shows the mean slope, $\theta$, of each facet over a tumbling period for the Apophis shape model for Case 15 projected onto an equirectangular projection. The slopes range from 0.4° to 45.3° with a median value of 13.4°. We note that the other cases produce slope maps that are qualitatively similar to Case 15. The slopes are shallower compared to the average slopes on the top-shaped and fast-spinning asteroid (101955) Bennu (17±2°) derived from high-resolution shape modeling from spacecraft data [Barnouin et al. 2019]. Though this small disparity could be due to a difference in resolution and quality of the two data sets used to produce the respective shape models (spacecraft imagery for Bennu vs. radar for Apohis). Thus, it may be the case that the slopes we model here are underestimated, as it is generally known that radar shapes tend to be "smoother" than reality. For example, the slope distribution obtained from radar data of Bennu [Nolan et al. 2013] show that a very small fraction of the modeled surface has slopes ≥ 30°, whereas these regions are prevalent in high-resolution shape models obtained from OSIRIS-REx data of Bennu [Barnouin et al. 2019, Daly et al. 2020].

High-slope regions are concentrated at latitudes < 60° and appear as extended hemisphere-crossing regions that also exhibit some symmetry about a prime meridian that we define here, which is found on the "head" region. We highlight these high-slope regions as they are the most likely locations to exhibit tidally induced surface mobilization, regardless of the exact mechanism. There are extended low-slope regions, particularly in the western hemisphere, that could serve as a catchment of destabilized regolith from surrounding regions. This particular low-slope region may resemble the fine-grained lowland regions of Itokawa [Fujiwara et al. 2006] and could thus serve as a safe landing location for a potential instrument, such as a seismometer [DellaGiustina,

Ballouz et al. 2024, Murdoch et al. 2024 T-5y], or the location of regolith-excavation experiments [DellaGiustina et al. 2023, Ballouz et al. 2023 LPSC, Barnouin et al. 2024a T-5y].

In Fig. 6b, we show the amplitude of the change in slope, $\delta\theta$, over a tumbling period. The changes in slope range from 0.13° to 0.86° with a median value of 0.57°. The $\delta\theta$ near the prime meridian ("head") and 180° longitude regions are small relative to the mid-longitude ("neck") regions. $\delta\theta$ also appears to exhibit symmetry about the prime meridian. The range in magnitude of $\delta\theta$ is smaller than those found in the study of Kim et al. (2023) (up to 1.5°), who included the effect of Earth tides on $\delta\theta$ and analyzed surface mobility due to the instantaneous change in spin period. While smaller, the $\delta\theta$ we find here occurs every tumbling period (tens of hours) rather than only once (during the Earth encounter) as modeled in Kim et al. (2023). Thus, the accumulated effect of many tumbling periods should induce larger changes over a longer timescale.

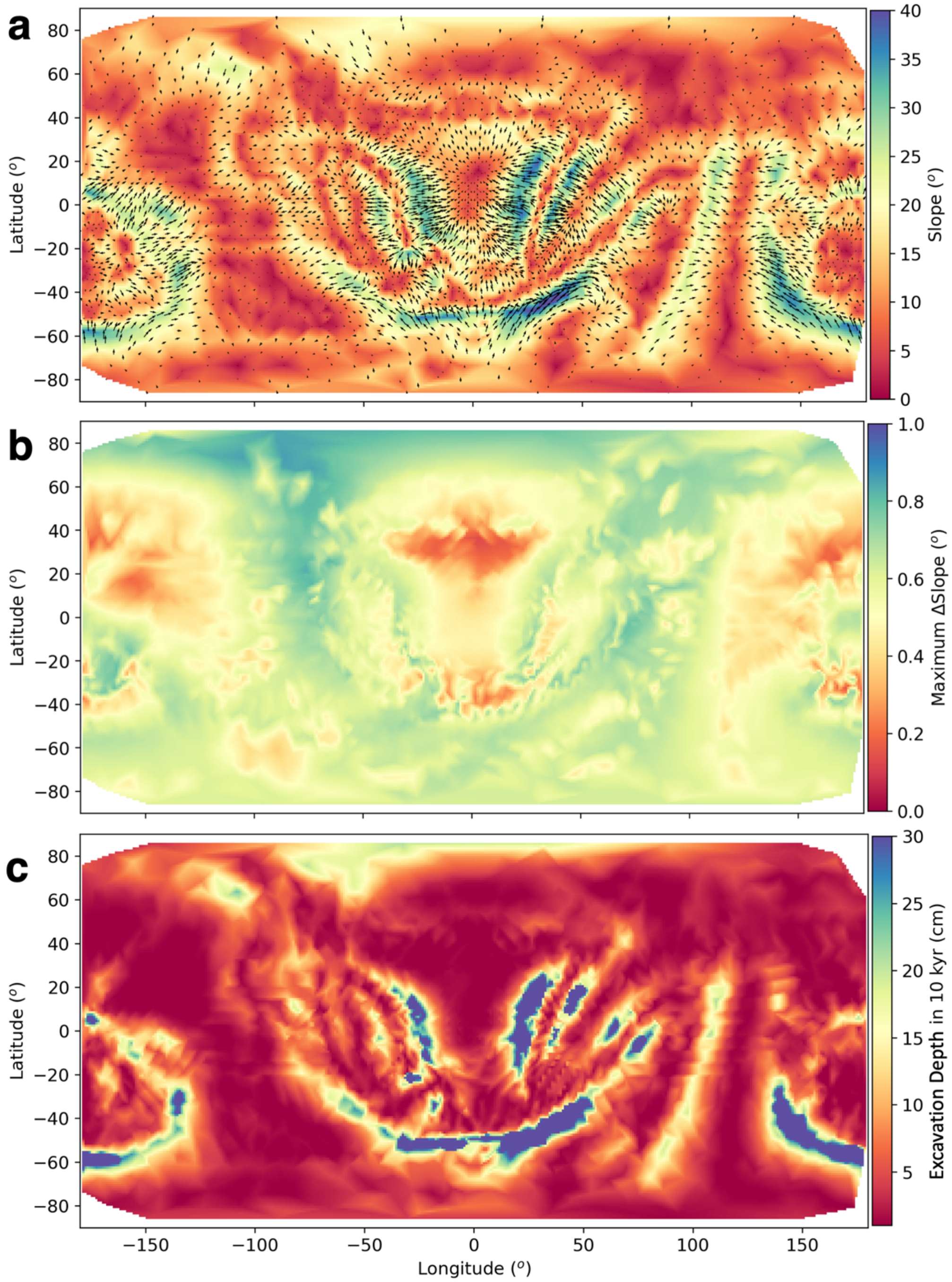

**Figure 6. a,** a map of the mean slope over a tumbling period for each facet of the Apophis shape model for Case 15 projected on to an equatorial map. The downslope direction is indicated by arrows whose length scales with the slope magnitude. **b,** a map of the amplitude of the slope change, $\delta\theta$, across Apophis for Case 15. **c,** a map of the excavation depth due to volume flux of cm-size particles over the course of 10 kyr due to periodic surface slope changes caused by the asteroid's tumbling state. This map is constructed using the granular mass flow prescription discussed in Sec. 5.1.

## 5. Long-Term Effects: Surface mobilization from tumbling

### *5.1 Long-Term Effects: Modeling granular flows through periodic tilting of granular beds*

To obtain an estimate for the magnitude of mass transport due to tumbling, we consider analytical theory of erosion for transport-limited downslope flow [Culling 1960]. For an asteroid surface with loose regolith with little to no cohesion, as determined directly for the surfaces of Bennu and Ryugu [Walsh & Ballouz et al. 2022, Arakawa et al. 2020] and inferred for Dimorphos based on the DART impact outcome [Barnouin et al. 2024b], the flow of grains is controlled by the transportation rate rather than the regolith supply or production rate. The flow rate, $q$, is typically directly proportional to the local surface slope, $\theta$. For higher slopes, fast granular flows are better modelled as a non-linear function that takes the angle of repose, $\theta_R$, into consideration. Based on studies of hill-slope failure in terrestrial and asteroid settings [Roering et al. 1999, Richardson et al. 2005], Ballouz et al. (2019) introduced an empirical formulation that captured the downslope flux of regolith, $Q$, due to time-varying slopes, $\delta\theta$, on the surface of the Martian moon Phobos: $Q = K \tan(\theta) \tan(\delta\theta) / \tan(\theta_R)$, where $K$ is the downslope flow constant in unit of volume flux per unit time. Ballouz et al. (2019) measured the value of $K$ by performing local simulations of a periodically tilting bed of regolith using *pkdgrav.* In those simulations, physical displacements of surface particles occur on steep slopes, which could be below the angle of repose, driven by the periodic slope changes due to Euler accelerations. Particles that are mobilized can topple and roll until coming to rest in the down-slope direction. The extent of particle creep was found to be a function of both $\theta$ *and* $\delta\theta$. For those simulations, the regolith was composed of a uniform size distribution of 1.5 mm-radius particles that had $\theta_R = 34°$. Here, we update the formulation of an empirical mass transport equation by considering sediment transport through dry ravel. Dry ravel is the geophysical process of particles rolling, bouncing, and sliding down a slope and is a dominant sediment transport process in steep arid and semi-arid landscapes [Gabet 2003]. For dry ravel, the downslope mass flux, $Q$, is given by:

$$Q = \frac{K}{\tan(\theta_R)\cos(\theta) - \sin(\theta)} \quad (1).$$

We modify this equation to incorporate the effects of a time-varying slope, $\delta\theta$, and slope, $\theta$, formulate a prescription for surface mass transport on asteroids that tumble, $Q_t$:

$$Q_t = K_t \frac{\tan^2(\delta\theta)\sin(\theta)}{\tan(\theta_R)\cos(\theta) - \sin(\theta)} \quad (2).$$

The modification of Eq. (1) is made from the analysis of experimental results (and the expectation) that the flow rate is directionally proportional to $\delta\theta$ and $\theta$. We also introduce a downslope flow constant specific to periodic tilting, $K_t$, in order to disambiguate with the constant in Eq. 1. We then use the periodic bed-tilting simulation results of Ballouz et al. (2019) and re-fit the data using Eq. (2) with a least-squares linear regression to find a best-fit value for $K_t = 479$ particles/m$^2$/period (Fig. 7a). This provides a prescription for the mass-volume flux for a given tumbling period, mean regolith particle size, and $\theta_R$. As shown in Fig. 6a, the new prescription shown in Eq. (2) describes the data well, with a square of the correlation coefficient, $r^2 = 0.91$. We prefer this new formulation as it is based on a previously established geophysical model for dry flows (Gabet et al. 2003) and does not require the piecewise linear solution of Ballouz et al. (2019).

We note that the simulations of Ballouz et al. (2019) were done under Phobos gravity and for particles with a mean radius $R_p$ = 1.5 mm. We use those results here as the simulations are computationally expensive. However, Phobos gravity is approximately two orders of magnitude larger than that of Apophis. There may be some dependence on volume flux due to periodic tilting with gravity, $g$. Therefore, to establish the applicability of these simulation-derived mass-flow prescriptions for Apophis, we consider a dimensionless Froude number, $Fr$, defined by $g$, $R_p$, and the periodic forcing frequency, $\omega_f$:

$$Fr = \frac{R_p \omega_f^2}{g}, \quad (3),$$

where $\omega_f$ would be inversely proportional to the orbital period for the Phobos case (7.65 h), as that sets the periodic change in Phobos' surface slopes, and the post-encounter tumbling period for the Apophis case (on the order of 100 h). The use of Froude number scaling is motivated by other works who have demonstrated its validity for understanding granular flow across different gravity regimes (e.g., Quillen et al. 2019 and references therein, Sunday et al. 2022, Tang et al. 2024). For seismic shaking, a Froude number with respect to the amplitude of shaking is typically defined [e.g., Matsumura et al. 2014]. Here, we adopt a scaling with $R_p$ as numerical studies have shown that the average size of particles in a shaken granular medium controls the onset of particle mobilization [e.g., Matsumura et al. 2014, Tang et al. 2022].

Using Eq. (3), we find an equivalent $Fr$ for $R_p$ = 1.5 mm and 5 mm for Phobos and Apophis, respectively. This means that surface motion driven by periodic tilting on Apophis reaches dynamic similitude to the Phobos simulations, for particle radii of 5 mm. For larger particle sizes, the Apophis scenario would have a larger value of $Fr$; thus, gravity is more easily overcome. We do not yet know the average particle size on Apophis; however, given the general observation of average particle sizes on smaller asteroids [e.g., DellaGiustina et al. 2019a], Apophis' regolith could be dominated by particles with $R_p > 5$ mm. In that scenario, Apophis regolith is more easily mobilized than Phobos regolith. Here, we conservatively consider that regolith mobilization driven by periodic surface tilting on Apophis is dynamically similar to that on Phobos.

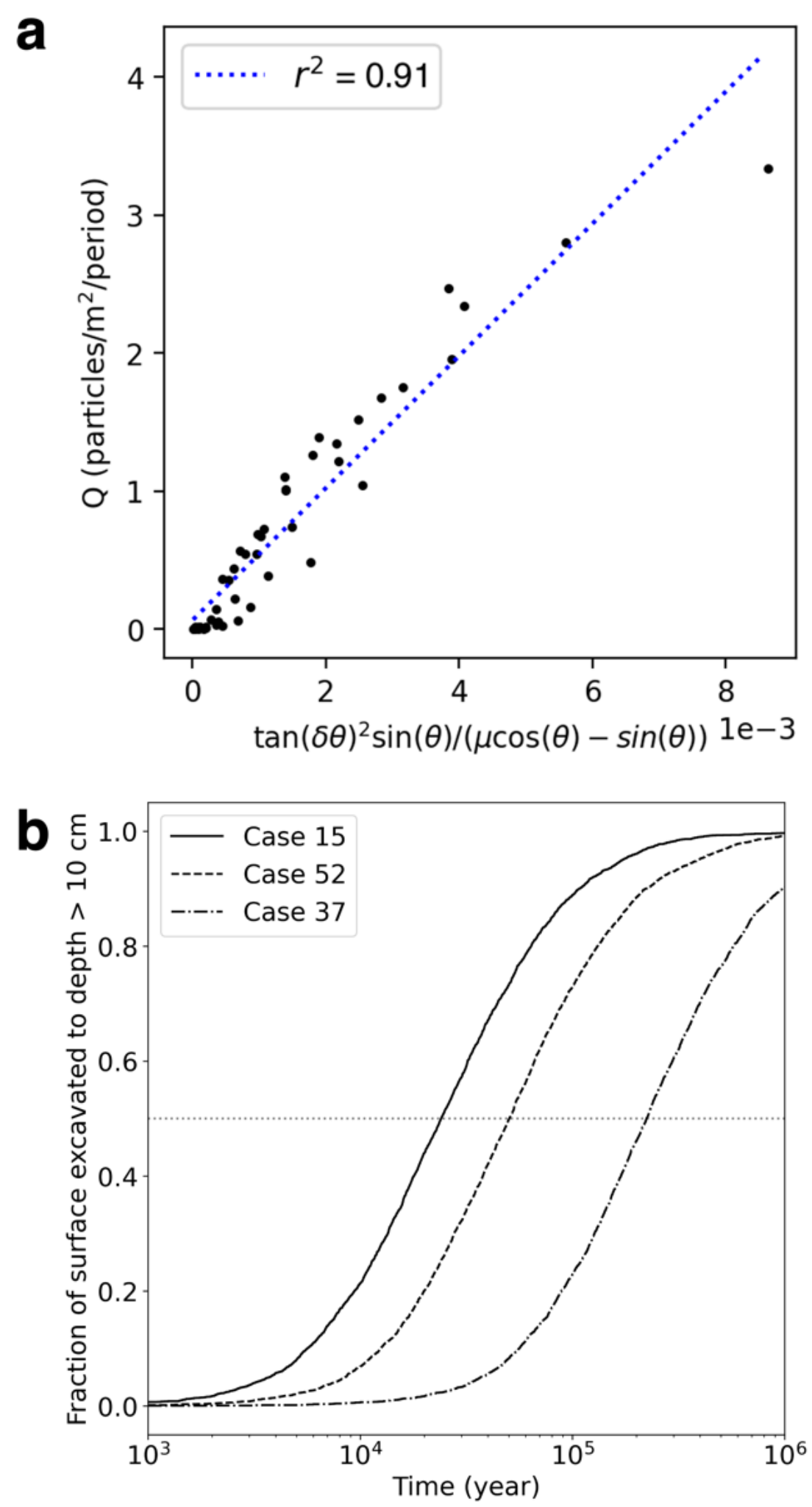

**Figure 7. a,** Mass movement results from periodic bed-tilting simulations of a granular bed [Ballouz et al. 2019a] are used to construct a new formulation based on the geophysical process of dry ravel (Eq. 2). **b,** The fraction of the Apophis surface "excavated" after a given time for three different cases (see Table A3). The horizontal dotted line denotes when 50% of the Apophis surface has been excavated to a depth > 10 cm. The 50%-excavation horizontal line crosses Case 15 (solid curve), Case 52 (dashed curve), and Case 37 (dot-dashed curve) at times of 24, 50, and 223 kyr, respectively.

*5.2 Long-Term Effects: The timescale for surface refreshing through tumbling for Apophis*

As we have established the feasibility of using Eq. (2) for Apophis, we model the volume flux from periodic tilting on a tumbling Apophis for different possible outcomes of its 2029 encounter as described in Section 3. For each facet of the radar shape model, we calculate a volume flux of 1 cm-radius particles over 1 tumbling period by computing the facet's $\theta$ and $\delta\theta$ values and assuming that Apophis regolith has $\theta_R = 40°$, which is equivalent to the largest slopes seen on small NEAs visited by spacecraft [e.g., Barnouin et al. 2019]. Using the area of each facet, we can then calculate the excavation depth from periodic tilting as a function of time. Here, we do not compute the deposition of regolith onto nearby facets, but simply attempt to get an order-of-magnitude estimate for how much volume could be mobilized from each facet. Computing the former would require a more sophisticated 2d model of mass flow on an asteroid surface, which is outside the scope of this study. As an example, Fig. 6c shows a map of the depth of excavation for Case 15 ($P_z$, $P_y$, and $P_x$ = 27.6 h, 32.5 h, and 493.4 h, respectively) after 10 kyr of tumbling at the same spin state as that following the 2029 encounter. The regions of high depth excavation coincide mostly with high-slope regions, though this is not always the case. As seen in Eq. (2), mass movement through this tumbling effect is strongest when both $\theta$ and $\delta\theta$ have large values. This unique combination of requirements allows for the distinguishing of periodic forcing effects on tumbling asteroids from other mechanisms that can also cause downslope force destabilization, such as seismic shaking.

Next, we consider the efficiency of the tumbling process at exposing fresh regolith on asteroids. However, we first must consider what could be the excavation depth and timescale needed to ensure that fresh regolith is exposed. The excavation depth requirement to reveal "fresh" regolith may be as small as exposing the underside of a surface particle, as modeled by Kim et al. (2023). Analysis of space-weathered rims on Itokawa samples and remote sensing data suggest that the depth may be smaller than 1 micrometer [Noguchi et al. 2014] and that the space-weathered rims may form on timescales as short as $10^2$–$10^4$ yr [Jin & Ishiguro 2022]. However, studies on space weathering of S-complex asteroid dynamical families reveal that space-weathering rates for these bodies may be as large as $10^6$ yr [Vernazza et al. 2008]. The regolith maturity parameter (Is/FeO ratio, i.e., the relative concentration of nanophase metallic iron, Is, to the total iron content, FeO) of twelve Apollo core samples show that the lunar regolith typically transitions between mature, weathered material to relatively fresh material around 20–50 cm below the surface [Lucey et al. 2006]. Though, we expect NEA regolith to experience far less extensive impact-based gardening at depth compared to lunar regolith due to the much smaller gravity on NEAs, which leads to less extensive secondary cratering. We proceed with a conservative approach to ascertain the time needed to excavate 10 cm of asteroid regolith.

Fig. 7b shows the result of our calculation for the fraction of the Apophis surface that is excavated to a depth greater than 10 cm as a function of time for three different cases: Case 15, Case 52, and Case 37. These cases were selected as they span a wide range of tumbling outcomes, with Case 15 having a similar magnitude in the body's spin rate about two axes, Case 52 showing a much larger spin rate about the minor axis ($P$ = 19.6 h) compared to the spin rate about the major axis ($P$ = 67.5 h), and Case 37 showing the smallest difference in spin rate about the major axis ($P$ = 54.7 h) and intermediate axis ($P$ = 78.2 h). We do not expect that regolith mobilization through tumbling can lead to 100% of the surface being refreshed, unless the space weathering depth is truly only "skin deep", as the weathered regolith is deposited into low elevation regions where it remains trapped. Thus, the model calculation in Fig.7b artificially shows eventual complete surface refreshing. For this model, we consider a surface to be refreshed if the majority of its surface is excavated to depths greater than 10 cm. Based on that criterion, the timescale required for surface refreshing for these cases is 24, 50, and 223 kyr for Case 15, 52, and 37, respectively.

These timescales are smaller than the characteristic time for rotational energy dissipation to return Apophis to principal-axis rotation, $\tau_{\mathrm{rot}}$ [Burns & Safronov 1973, Richardson et al. 1998]:

$$\tau_{rot} = \frac{MQ_{tidal}}{\rho_{bulk} S_3^2 R_{ast}^2 \omega^3} \quad (4),$$

where $M$ is the asteroid's rigidity, $Q_{tidal}$ is its quality factor, $R_{ast}$ is the asteroid radius, and ω is its spin frequency. $S_3^2$ is a shape factor ranging between 0.01 for nearly spherical bodies and $0.1H^2$ for non-spherical bodies with oblateness, $H = (I_3 - I_1)/I_3$, where $I_1$ and $I_3$ are the major and minor principal moments of inertia, respectively. Using the radar model of Brozovic et al. (2018), Apophis has $H = 0.27$. On the basis of tidal evolution calculations for NEA binaries, Taylor & Margot (2011) estimated the value of $MQ_{tidal}$ to range between ~ $6\times10^6$ and $1\times10^{10}$ N m$^{-2}$. Recently, DellaGiustina & Ballouz et al. (2024) calculated a value of $MQ_{tidal} = 2\times10^8$ N m$^{-2}$ using numerical simulations of rubble piles in a system that resembles that of (65803) Didymos, which is within the range of estimates from Taylor & Margot (2011). For a tumbling Apophis with an effective rotation period of 25 h, $\rho_{\rm bulk} = 2.2$ g/cm$^3$, $R_{ast} = 180$ m, and $MQ_{tidal} = 2\times10^8$ N m$^{-2}$, Eq. (4) gives $\tau_{\rm rot} = 35$ Myr. Using the lower end of $MQ_{tidal}$ from Taylor & Margot (2011), we calculate a lower bound for $\tau_{\rm rot} = 1.05$ Myr. Therefore, a tumbling Apophis could indeed be resurfaced through our proposed mechanism before its rotational state is dampened.

We note that significant mass motion could lead to changes in the moment of inertia and dissipate energy through friction. These effects could lead to faster damping timescales. If that timescale is still smaller than the space weathering timescale, then this could explain why the majority of Q-class asteroids are found in principal axis rotation, even if resurfacing was originally caused by tumbling. Alternatively, large scale mass motion that could contribute to rotational damping would also ostensibly refresh the surface. In essence, the rate of rotational damping would be directly related to, and caused by, surface refreshing.

**6. Discussion**

In this work, we have investigated how close encounters of NEAs with terrestrial planets can drive two physical mechanisms that may cause surface changes: 1) short-timescale seismic shaking and 2) long-timescale tumbling-induced downslope motion. We investigate this mechanism through the lens of the Apophis close approach with Earth on April 13, 2029, but our results are applicable to the general problem of NEA surface refreshing.

In short timescales, our modeling shows that seismic events initiate approximately an hour before perigee and continue past the encounter as the asteroid's interior settles into a post-encounter equilibrium. The SSDEM simulations show global surface shaking that can reach acceleration values of $\Gamma = |a/g| = 1$, for the material parameters adopted in these simulations. We note that the magnitude of seismic acceleration due to tides will vary with the magnitude of the spring constants ($k_n$ and $k_t$) used in the simulations, which define the stiffness of the rubble-pile interior. Therefore, we expect that differences in the internal structure and stiffness of real asteroids could be measured by seismometers [DellaGiustina & Ballouz et al. 2024].

Previous studies on seismic shaking on asteroids have shown that particle lofting, downslope mass flow, and granular convection may occur when $\Gamma = 1$ [Matsumura et al. 2014, Maurel et al. 2017, Tancredi et al. 2016, Perera et al. 2016, Quillen et al. 2019, Tang et al. 2023, 2024]. It may be possible that similar phenomena could be seen during Apophis' closest encounter with Earth. However, those studies model shaking amplitudes that are close to the mean particle size in the granular medium. Here, our simulations show that the shaking has high frequencies (>0.1 Hz), and the shaking amplitudes required to achieve $\Gamma = 1$ would be <0.1 mm. Furthermore, if part of the asteroid surface experiences seismic accelerations larger than local gravity that does not necessarily translate to ground velocities that are would be sufficient to launch particles to escaping trajectories [Garcia et al. 2015]. Thus, the close passage of Apophis may be able to loft dust particles, if they exist on Apophis; however, it is unclear if the modeled shaking could cause widespread mass mobilization and boulder lofting. Any lofted boulders or dust would likely re-impact or escape from Apophis [Chesley et al. 2020] by the time that OSIRIS-APEX arrives [DellaGiustina et al. 2023]. Large-scale mobilization may produce distinct re-accretion patterns on the asteroid or place particles on orbits that are stable for tens of years [Valvano et al. 2022] and that would be observable by OSIRIS-APEX. Further laboratory and computational research into the feasibility of driving down-slope mass motion and particle lofting through low-amplitude vibrations in low gravity would provide clarity on the outcome of the Apophis close encounter and tidally-driven surface refreshing of asteroids.

Regardless of whether tidally induced seismic shaking can cause surface refreshing on Apophis, our simulations clearly show that tides would generate quaking events that would be measurable with modern seismometers (Fig. 5). Such seismometers are on the path to be sufficiently mature for fielding on potential Apophis missions that could explore the asteroid before its close approach with Earth in 2029 [Murdoch et al. 2024 T-5y, DellaGiustina & Ballouz et al. 2024].

Over long timescales, our modeling shows that tidally induced rotation-state changes can cause Apophis to tumble at rates that would lead to periodic changes in slope of up to ~1°. This periodic change in surface slopes can cause a gradual creep-like motion on the surface that can cause regional resurfacing in timescales of ~10–100 kyr, depending on the rotation state. The rate of creep-induced mass movement for tumblers is slightly faster than that observed on non-tumbling rubble-pile asteroid Bennu [Jawin et al. 2020]. The wide range in surface refreshing timescales for tumbling is reflective of the differences in the range of $\delta\theta$ across the different cases. Cases that result in similar spin rates about two axes lead to larger $\delta\theta$ across the asteroid surface compared to a case that has a faster spin rate about one of its axes. Put another way, a body that is in a faster tumbling state will experience faster surface refreshing. Thus, no two tumblers are created "equal" in terms of their potential for surface refreshing. For the case of Apophis, OSIRIS-APEX may detect surface changes after it arrives at Apophis if its tumbling state is similar to that shown in Case 15. In particular, the mission may detect down-slope creep of meter-scale boulders stranded on the surface, or the gradual refreshing of its surface if the space weathering depth on S-classes is indeed only "skin deep" [Noguchi et al. 2014, Jin & Ishiguro 2022]. In contrast to the short-term seismic effects, long-term tumbling induced creep would be localized to regions with large initial slopes. Detailed observations of these regions by OSIRIS-APEX would allow for the distinguishing of long-term and short-term effects. Furthermore, characterization of the rigidity and quality factor of Apophis with an *in situ* seismometer would provide insight into the long-term evolution of the rotation state and surface changes of Apophis through tumbling.

Some tumblers may even experience surface mobilization at a rate that is slower than the space weathering rate. This range of outcomes is consistent with the observation that tumbling asteroids are found across the S-complex (as shown in Section 2) and the observation that not all asteroids that could have had close encounters with Earth are Q-classes (see Fig. 2b of Binzel et al. 2010). Thus, the work here presents a potential solution to the "resurfacing paradox": S-complex asteroids that experience distant planetary encounters (up to 15 planetary radii) are resurfaced when tidal torques induce a change in the rotational state that can drive long-term changes through tumbling. Non-disruptive planetary encounters may also induce short-term changes through quaking driven by subtle internal reconfigurations of rubble-pile asteroids that lead to seismic accelerations that can reach levels equivalent to the asteroid's gravitational acceleration (see Fig. A3). Further observational studies that characterize the spin states of known

S-complex asteroids and/or characterize the spectral properties of known tumblers would provide insight into the timescale for surface refreshing through these mechanisms.

While the shaking and tumbling effects we describe in this work may not be the only mechanism that could produce surface changes induced by tides, it does provide a clear testable hypothesis: small bodies in strong tidal environments should experience some combination of seismic shaking and gradual surface movement through periodic surface slope changes. As previously highlighted, a similar mechanism may be operating on Phobos, which will be closely observed by the MMX mission [Ballouz et al. 2019a]. Furthermore, work by [Agrusa et al. 2022] has shown that a similar mechanism of surface refreshing through non-principal axis rotation may be operating on Dimorphos following the DART impact. If so, the Hera mission may possibly observe mass movement on Dimorphos during its mission [Michel et al. 2022]. Finally, this work provides a basis for understanding the full extent of surface refreshing on Apophis through shaking and tumbling that will be tested through the detailed observation campaign of the OSIRIS-APEX mission [DellaGiustina et al. 2023].

**Acknowledgements**

R.-L.B. was supported in part by internal funds from the Johns Hopkins University Applied Physics Laboratory. R.-L.B., K.J.W., D.N.D., V.J.B., and A.M. acknowledge funding from the NASA MATISSE program through grant 80NSSC23K0174. D.N.D. was supported by NASA contract No. NNM10AA11C. H.A. was supported by the French government, through the UCA J.E.D.I. Investments in the Future project managed by the National Research Agency (ANR) with the reference number ANR-15-IDEX-01. P.M. acknowledges funding support from CNES, ESA and The University of Tokyo. N.M. acknowledges support from CNES. J.V.D. and D.C.R. acknowledge funding from the NASA SSERVI program through grant 80NSSC19M0216 (GEODES). We thanks the anonymous reviewers for their detailed feedback on the original versions of this manuscript.

## Appendix

*A.1 Discrete-element simulations with pkdgrav*

*pkdgrav* is a combined *N*-body gravity and discrete-element method (DEM) collisional code capable of accurately simulating the complexity of grain-grain and grain-boundary interactions through a soft-sphere DEM (SSDEM) [Richardson, 2000, Schwartz et al. 2012, Ballouz 2017]. In SSDEM, collisions of spherical grains are resolved by allowing them to slightly overlap and then applying multi-contact and multi-frictional forces, including static, rolling and twisting friction. Modeling grain friction accurately is a critical component for high-fidelity granular physics simulations. For *pkdgrav*, rolling friction and inter-particle cohesion models have been implemented that allow a more accurate modeling of grain shape, angularity, and inter-particle cohesion. In this manner, the code accurately simulates the complexity in the interaction of irregularly shaped grains by capturing their bulk behavior correctly (despite modeling spherical particles). New rolling friction and inter-particle cohesion models have recently been implemented [Zhang et al. 2017; 2018] and allow for more accurate modeling of grain shape, angularity, and electrostatic interactions. The parameters used in this study are tabulated in Table A2 and are chosen to simulate rubble piles with an angle of repose of ~35°.

In this study, we choose to ignore the effects of inter-particle cohesion. Spacecraft-regolith interactions on small C-complex NEAs have shown that their surfaces have very low levels of cohesion [Arakawa et al. 2020, Walsh & Ballouz et al. 2022]. Recent observations from the DART mission have suggested that this is also true for NEAs in the S-complex [Barnouin et al. 2024]. However, observations of terrace features on Bennu's high latitudes suggest that even low levels of cohesion may control the formation of geophysical features [Barnouin et al. 2022]. Observations of similar features and/or lineaments on Apophis will help to inform the extent that inter-particle cohesion plays a role in influencing surface changes from planetary encounters.

In order to adequately resolve collisions, we set the normal and tangential SSDEM spring constants to be $1.58 \times 10^9$ kg/s$^2$ and $4.5 \times 10^8$ kg/s$^2$, respectively. These values correspond to elastic moduli $Es$ = 33 to 100 MPa and longitudinal seismic wave speeds $v_p$ = 78 to 134 m/s for the simulation's 5 to 15 m radius particles, respectively. The elastic moduli are estimated by considering the DEM spring constant, $k_n$, and the radius of the particle, $R$, such that $Es = k_n/(\pi R)$. Elastic moduli and seismic wave speeds of meteorites are typically on the order of 10 GPa and 2.5 km/s, respectively [Cotto-Figueroa et al. 2016]. Here, the spring constants are chosen such that each particle of the simulated asteroid pile could be considered a representative volume element of a granular pile (smaller dust, pebble, cobbles, and boulders), which typically have bulk $Es \sim 10$ MPa and $v_p \sim 100$ m/s, respectively [e.g., Goddard 1990, Goldreich & Sari 2009]. [DellaGiustina & Ballouz et al. 2024] showed that the dissipative properties of rubble piles modeled with these material properties provide a reasonable match to constraints based on tidal dissipation from observations of asteroid binaries [Taylor & Margot 2011].

In our analysis of global seismicity in Section 3, surface particles are defined by first calculating the alpha shape of the centers of the simulated particles that have been shifted towards the COM by their radius (see Ballouz et al. 2019b for details on alpha shapes for rubble piles). Then, surface particles are defined as those that lie outside the alpha shape. Ground motion for each surface particle is calculated using the same scheme described in Section3 for a single particle.

*A.2 Modeling surface slopes for a tumbling asteroid*

In order to evaluate time-varying surface slopes of an asteroid, we consider the net acceleration on a surface element *i* at time *t*, $\boldsymbol{a}_{i,t}^{\text{net}}$, which is given by:

$$\boldsymbol{a}_{i,t}^{\text{net}} = \boldsymbol{a}_{i}^{\text{grav}} + \boldsymbol{a}_{i,t}^{\text{cent}} + \boldsymbol{a}_{i,t}^{\text{Euler}}, \qquad (A1)$$

where $\boldsymbol{a}_{i}^{\text{grav}}$, $\boldsymbol{a}_{i,t}^{\text{cent}}$, $\boldsymbol{a}_{i,t}^{\text{Euler}}$ are the gravitational, centrifugal, and Euler acceleration, respectively. The surface slope of element *i* at time *t*, $\theta_{i,t}$ is defined as:

$$\theta_{i,t} = \hat{\boldsymbol{n}}_i \cdot \hat{\boldsymbol{a}}_{i,t}^{\text{net}}, \qquad (A2)$$

where $\hat{\boldsymbol{n}}_i$ is a facet's surface normal and $\hat{\boldsymbol{a}}_{i,t}^{\text{net}}$ is the $\boldsymbol{a}_{i,t}^{\text{net}}$ unit vector. We use the Apophis radar shape model of Brozovic et al. (2018), which is composed of 3,996 facets that have an average length of approximately 15 m. To calculate $\boldsymbol{a}_{i}^{\text{grav}}$, we use the algorithm of Werner & Scheeres (1996) as implemented in Barnouin et al. (2020). The algorithm allows us to compute the geoid and gravitational acceleration at each of Apophis' surface facets. For these calculations, we assume Apophis has a homogeneous interior bulk density, $\rho_{\text{bulk}}$ = 2.2 g/cm$^3$. In order to evaluate the centrifugal and Euler accelerations, we integrate the rigid-body equations of motion for the torque-free case:

$$\begin{aligned} I_1\dot{\omega}_1 - \omega_2\omega_3(I_2 - I_3) &= 0 \\ I_2\dot{\omega}_2 - \omega_3\omega_1(I_3 - I_1) &= 0 \\ I_3\dot{\omega}_3 - \omega_1\omega_2(I_1 - I_2) &= 0, \end{aligned} \qquad (A3)$$

where $I_k$ are the principal moments of inertia of Apophis, $\omega_k$ are the spin components in the body frame, and $\dot{\omega}_k$ are the time derivatives of $\omega_k$ ($k$ = 1, 2, 3). The principal moments of inertia are calculated using the Apophis radar shape model, using the formulation of Eberly (2009). We find mass normalized $I_1$, $I_2$, and $I_3$ = 10013, 12987, and 13708 m$^2$, respectively. We numerically integrated Eq. (A3) using the Python package *SciPy* and its *odeint* method, which solves a system of differential equations using the *lsoda* integrator [Virtanen et al. 2020]. The centrifugal and Euler accelerations of surface element *i* are then calculated at a given point on the Apophis surface:

$$\boldsymbol{a}_{i,t}^{\text{cent}} = (\boldsymbol{\omega}_t \times \boldsymbol{r}_i) \times \boldsymbol{\omega}_t, \qquad (A4)$$

$$\boldsymbol{a}_{i,t}^{\text{Euler}} = \boldsymbol{r}_i \times \dot{\boldsymbol{\omega}}_t, \qquad (A5)$$

where $\boldsymbol{\omega}_t$ is the spin angular velocity vector at time $t$ and $\boldsymbol{r}_i$ is the position vector of the surface element. Solutions for Eqs. (A4–A5) at each time step are then used to evaluate $\theta_{i,t}$ via Eq. (A2).

| Parameter | Value |
| --- | --- |
| Normal Restitution coefficient, $\varepsilon_N$ | 0.55 |
| Tangential Restitution coefficient, $\varepsilon_T$ | 0.55 |
| Static friction coefficient, $\mu_S$ | 0.6 |
| Rolling friction coefficient, $\mu_R$ | 1.05 |
| Twisting friction coefficient, $\mu_T$ | 1.3 |
| Shape parameter, $\beta$ | 0.6 |
| Normal spring constant (kg $s^{-2}$), | $1.58 \times 10^9$ |
| Tangential spring constant (kg $s^{-2}$), | $4.5 \times 10^8$ |
| Timestep (s) | 0.01 |

**Table A1.** Summary of *pkdgrav* restitution and friction coefficients used in this study. See Zhang et al. (2017) for full details on particle collision modeling in the SSDEM implementation.

| Structure | $\rho_{bulk}$ (g/cm$^3$) | $\rho_{head}$ (g/cm$^3$) | COM-COF(m) |
| --- | --- | --- | --- |
| homogeneous | 2.2 | 2.2 | – |
| heterogeneous | 2.2 | 2.8 | 9.7 |
| heterogeneous | 2.2 | 3.3 | 17.8 |

**Table A2.** Summary of internal structures used for tidal encounter simulations. $\rho_{bulk}$: bulk density; $\rho_{head}$: bulk density of head; COM-COF: center-of-mass to center-of-figure offset.

| Case | $\rho_{head}$ (g/cm$^3$) | $\phi$ (°) | $P_{final}$ (h) | $P_x$ (h) | $P_y$ (h) | $P_z$ (h) |
|---|---|---|---|---|---|---|
| 1 | 2.2 | 7.9 | 24.8 | 63.5 | 114.9 | 27.7 |
| 2 | 2.2 | 12.5 | 29.4 | 65.9 | 92.0 | 35.2 |
| 3 | 2.2 | 26.2 | 24.4 | 160.7 | 37.6 | 32.7 |
| 4 | 2.2 | 28.9 | 28.4 | 105.4 | 47.3 | 37.8 |
| 5 | 2.2 | 32.0 | 34.2 | 76.1 | 74.6 | 44.7 |
| 6 | 2.2 | 42.5 | 32.9 | 93.0 | 67.3 | 41.3 |
| 7 | 2.2 | 52.5 | 35.2 | 93.2 | 58.2 | 50.1 |
| 8 | 2.2 | 60.7 | 34.3 | 103.4 | 108.0 | 38.7 |
| 9 | 2.2 | 73.1 | 30.7 | 113.0 | 45.1 | 45.1 |
| 10 | 2.2 | 80.3 | 30.2 | 142.0 | 195.5 | 31.3 |
| 11 | 2.2 | 93.1 | 25.5 | 133.1 | 38.2 | 35.4 |
| 12 | 2.2 | 99.9 | 24.7 | 360.4 | 393.4 | 24.8 |
| 13 | 2.2 | 112.0 | 22.2 | 182.9 | 34.4 | 29.4 |
| 14 | 2.2 | 118.7 | 21.5 | 266.3 | 461.8 | 21.6 |
| 15 | 2.2 | 129.3 | 21.0 | 493.4 | 32.5 | 27.6 |
| 16 | 2.2 | 136.9 | 20.6 | 99.5 | 255.6 | 21.2 |
| 17 | 2.2 | 144.3 | 21.8 | 479.3 | 33.3 | 28.9 |
| 18 | 2.2 | 154.7 | 21.8 | 70.8 | 156.8 | 23.2 |
| 19 | 2.8 | 7.9 | 24.2 | 57.8 | 89.8 | 27.9 |
| 20 | 2.8 | 12.5 | 29.8 | 63.2 | 80.8 | 37.2 |
| 21 | 2.8 | 26.2 | 24.3 | 186.6 | 38.8 | 31.5 |
| 22 | 2.8 | 28.9 | 29.4 | 112.8 | 50.2 | 38.4 |
| 23 | 2.8 | 32.0 | 36.1 | 78.9 | 70.8 | 49.6 |
| 24 | 2.8 | 42.5 | 35.5 | 98.2 | 75.2 | 44.1 |
| 25 | 2.8 | 52.5 | 37.7 | 103.5 | 57.8 | 56.6 |
| 26 | 2.8 | 60.7 | 37.5 | 115.1 | 125.9 | 41.8 |
| 27 | 2.8 | 73.1 | 31.5 | 129.1 | 44.8 | 47.2 |
| 28 | 2.8 | 80.3 | 31.4 | 176.3 | 209.2 | 32.3 |
| 29 | 2.8 | 93.1 | 25.0 | 143.0 | 38.8 | 33.6 |
| 30 | 2.8 | 99.9 | 24.1 | 1096.9 | 329.1 | 24.2 |
| 31 | 2.8 | 112.0 | 21.2 | 177.7 | 35.2 | 26.9 |
| 32 | 2.8 | 118.7 | 20.3 | 152.8 | 274.3 | 20.5 |
| 33 | 2.8 | 129.3 | 20.0 | 380.0 | 33.0 | 25.2 |
| 34 | 2.8 | 136.9 | 19.4 | 77.1 | 155.6 | 20.2 |
| 35 | 2.8 | 144.3 | 21.1 | 825.4 | 33.9 | 27.0 |
| 36 | 2.8 | 154.7 | 20.7 | 60.6 | 108.6 | 22.6 |
| 37 | 3.3 | 7.9 | 23.7 | 54.7 | 78.2 | 28.0 |
| 38 | 3.3 | 12.5 | 29.9 | 61.8 | 74.8 | 38.5 |
| 39 | 3.3 | 26.2 | 24.2 | 208.3 | 39.7 | 30.9 |
| 40 | 3.3 | 28.9 | 30.1 | 116.9 | 52.5 | 38.8 |
| 41 | 3.3 | 32.0 | 37.5 | 80.1 | 69.9 | 53.5 |
| 42 | 3.3 | 42.5 | 37.4 | 100.8 | 81.9 | 46.2 |
| 43 | 3.3 | 52.5 | 39.5 | 111.2 | 58.3 | 61.5 |
| 44 | 3.3 | 60.7 | 40.0 | 124.2 | 140.8 | 44.2 |
| 45 | 3.3 | 73.1 | 32.0 | 143.5 | 44.4 | 48.7 |
| 46 | 3.3 | 80.3 | 32.2 | 211.1 | 216.5 | 33.0 |
| 47 | 3.3 | 93.1 | 24.6 | 154.0 | 38.8 | 32.6 |
| 48 | 3.3 | 99.9 | 23.7 | 2437.1 | 289.6 | 23.8 |
| 49 | 3.3 | 112.0 | 20.6 | 175.7 | 35.7 | 25.5 |
| 50 | 3.3 | 118.7 | 19.5 | 116.2 | 206.7 | 19.8 |
| 51 | 3.3 | 129.3 | 19.4 | 323.7 | 33.5 | 23.9 |
| 52 | 3.3 | 136.9 | 18.6 | 67.5 | 118.5 | 19.6 |
| 53 | 3.3 | 144.3 | 20.7 | 1649.7 | 34.4 | 25.9 |
| 54 | 3.3 | 154.7 | 20.1 | 55.6 | 88.7 | 22.2 |

**Table A3.** Summary of outcomes for tidal encounter simulations. $\rho_{\text{head}}$: bulk density of head; $\phi$: angle between major axis and Earth-Apophis vector at perigee; $P_{\text{final}}$: post-encounter spin period of Apophis; $P_x$, $P_y$, and $P_z$: post-encounter spin period of Apophis about major, intermediate, and minor axes, respectively.

| Asteroid Designation | Tumbling Flag | Spectral Class | Spin Period (h) |
|---|---|---|---|
| 1997 AE12 | NT0 | Q | 1880.00 |
| 1998 SJ70 | T0 | Q | 19.15 |
| 1998 XB | NT0 | Q | 520.00 |
| 1999 BY9 | T? | Q | 700.00 |
| 1999 NC43 | T?A | Q | 34.49 |
| 2000 AC6 | T | Q | 2.44 |
| 2001 FM129 | T | Q | 38.56 |
| 2002 NY40 | T | Q | 19.98 |
| 2003 FH | T0 | Q | 13.94 |
| 2003 QO104 | T? | Q | 114.40 |
| 2003 UV11 | T | Q | 18.25 |
| 2005 ED318 | NT0 | Q | 17.16 |
| 2008 HS3 | T | Q | 10.68 |
| 2010 XZ67 | T? | Q | 15.04 |
| 2012 QG42 | T? | Q | 24.22 |
| 2014 QK434 | T0 | Q | 78.40 |
| 1992 KD | T0 | Q | 226.40 |
| 1991 RC | T? | Q | 23.60 |
| 1989 VB | T0 | Sq | 14.53 |
| 1998 OH | T? | Sq | 2.58 |
| 2000 FO10 | T? | Sq | 53.88 |
| 2000 NF5 | T | Sq | 59.26 |
| 2000 QW7 | T | Sq | 71.57 |
| 2000 XK44 | T | Sq | 51.90 |
| 2001 HG31 | T | Sq | 60.61 |
| 2001 JV1 | RNT0 | Sq | 29.00 |
| 2001 US16 | T | Sq | 14.39 |
| 2005 FC3 | T0 | Sq | 430.00 |
| 2005 GN59 | T- | Sq | 38.69 |
| 2005 UL5 | T? | Sq | 3.46 |
| 2007 PA8 | T | Sq | 102.24 |
| 2007 TQ24 | T? | Sq | 52.30 |
| 2007 TU24 | T | Sq | 26.00 |
| 2014 AD17 | T- | Sq | 8.48 |
| 2004 MN4 | T | Sq | 30.56 |
| 1998 SF36 | T- | Sq | 12.13 |
| 1981 QA | T- | Sq | 149.40 |
| 1987 SB | RT0 | Sq | 67.50 |
| 1988 TJ1 | T | Sq | 21.10 |
| 1987 SY | T | Sq | 56.48 |
| 1982 DV | T | Sq | 75.00 |
| 1989 AC | T | Sq | 176.00 |
| 1989 UR | NT0 | S | 73.00 |
| 1999 FA | T- | S | 10.09 |
| 2002 TD60 | T+ | S | 2.85 |
| 2002 AG29 | T0 | S | 19.64 |
| 2000 GQ146 | NT0 | S | 51.00 |
| 1994 PC | T0 | S | 35.90 |
| 1988 EF | T- | S | 89.30 |
| 1999 FK21 | T | S | 28.08 |
| 1997 BR | T | S | 33.64 |
| 2000 AP59 | T | S | 64.00 |
| 1994 TW1 | T? | S | 97.10 |
| 1987 SF7 | T? | S | 453.00 |
| 2001 QP181 | T0 | S | 17.05 |
| 2011 GP59 | T+ | S | 0.12 |
| 2007 PU11 | T0 | S | 56.80 |
| 2000 ED104 | NT0 | S | 43.00 |
| 1993 OM7 | T0 | S | 26.00 |
| 1989 FB | T0 | S | 37.65 |
| 2003 SD220 | T? | S | 285.00 |
| 2018 RC | T | S | 0.17 |
| 2006 UM216 | T | S | 31.73 |
| 1998 YP11 | T- | S | 38.60 |
| 2003 RB | T? | S | 37.50 |
| 1984 QY1 | T | S | 45.50 |
| 2010 LJ14 | T0 | S | 113.00 |
| 2001 XY10 | T0 | S | 43.50 |
| 2002 TD66 | T0 | S | 9.46 |
| 2001 WG2 | T | S | 46.08 |
| 2001 VS78 | T- | S | 40.55 |
| 1981 WO1 | T0 | S | 280.00 |
| 2001 WC47 | T0 | S | 16.51 |

**Table A4.** List of potential tumbling S-complex asteroids in the MITHNEOS database. See [Warner et al. 2021] for details on tumbling flags.

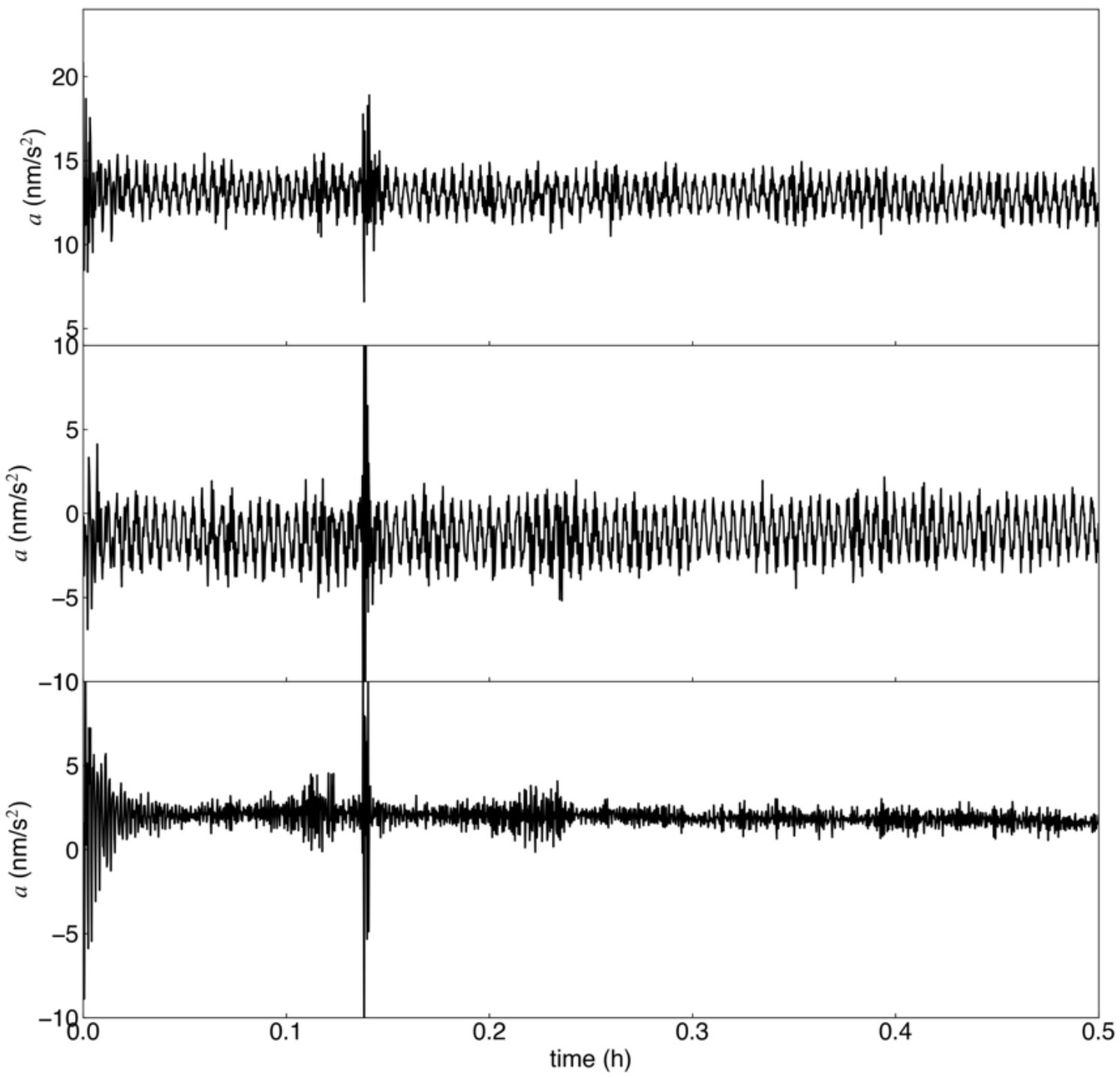

**Figure A1.** The radial acceleration for the surface particles highlighted in Fig. 3a (*top panel*), Fig. 3b (*middle panel*), and Fig. 3c (*bottom panel*), after the initial settling period. Background vibration levels are on the order of a 5-15 nm/s$^2$, which is three-to-four orders of magnitude smaller than the seismic accelerations induced by the close encounter.

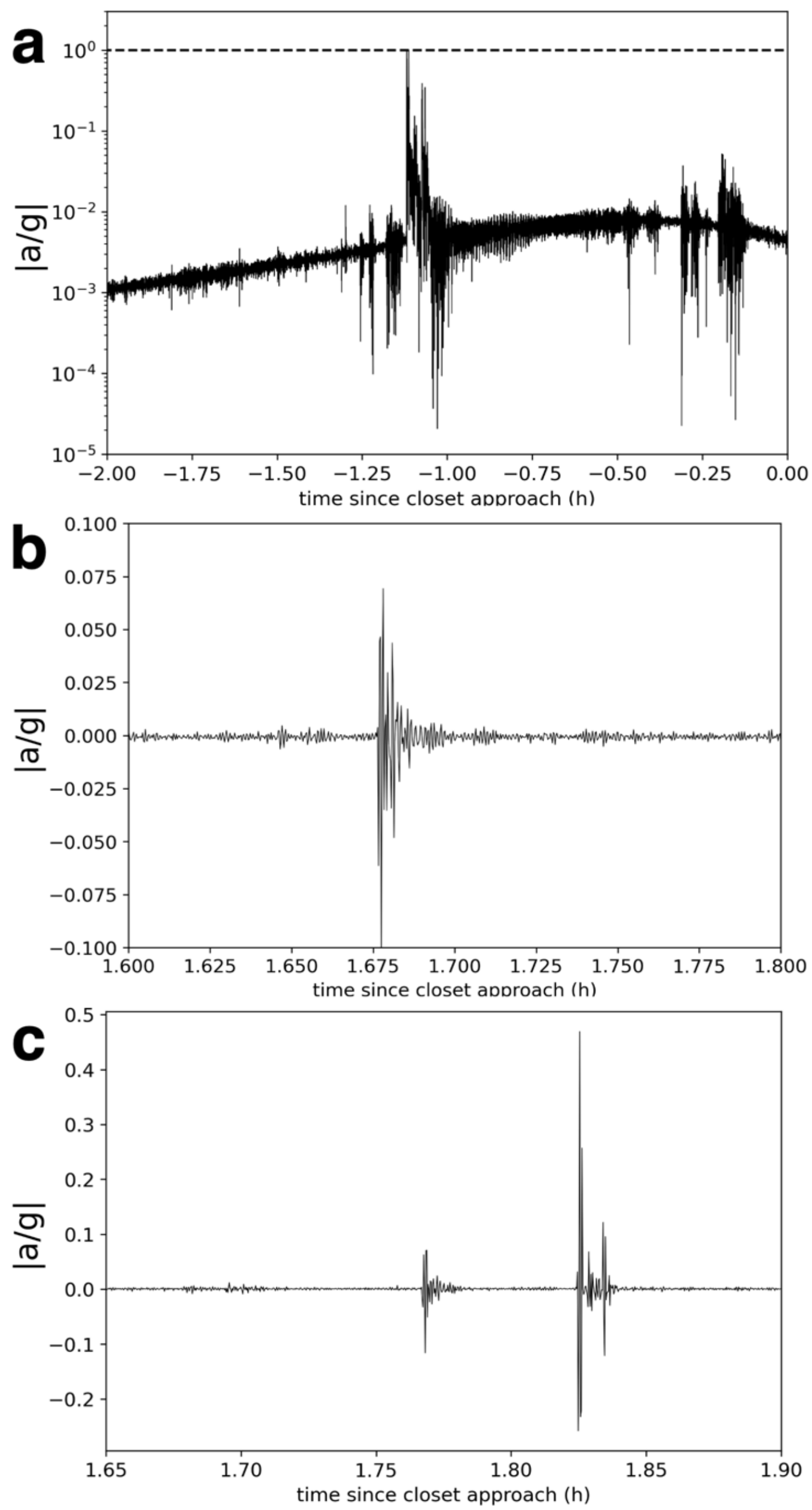

**Figure A2.** Closeups of the simulated ground motion on Apophis shown in Fig. 3 for the events in the cyan-shaded region (**a**), the magenta-shaded region (**b**), and the yellow-shaded region (**c**).

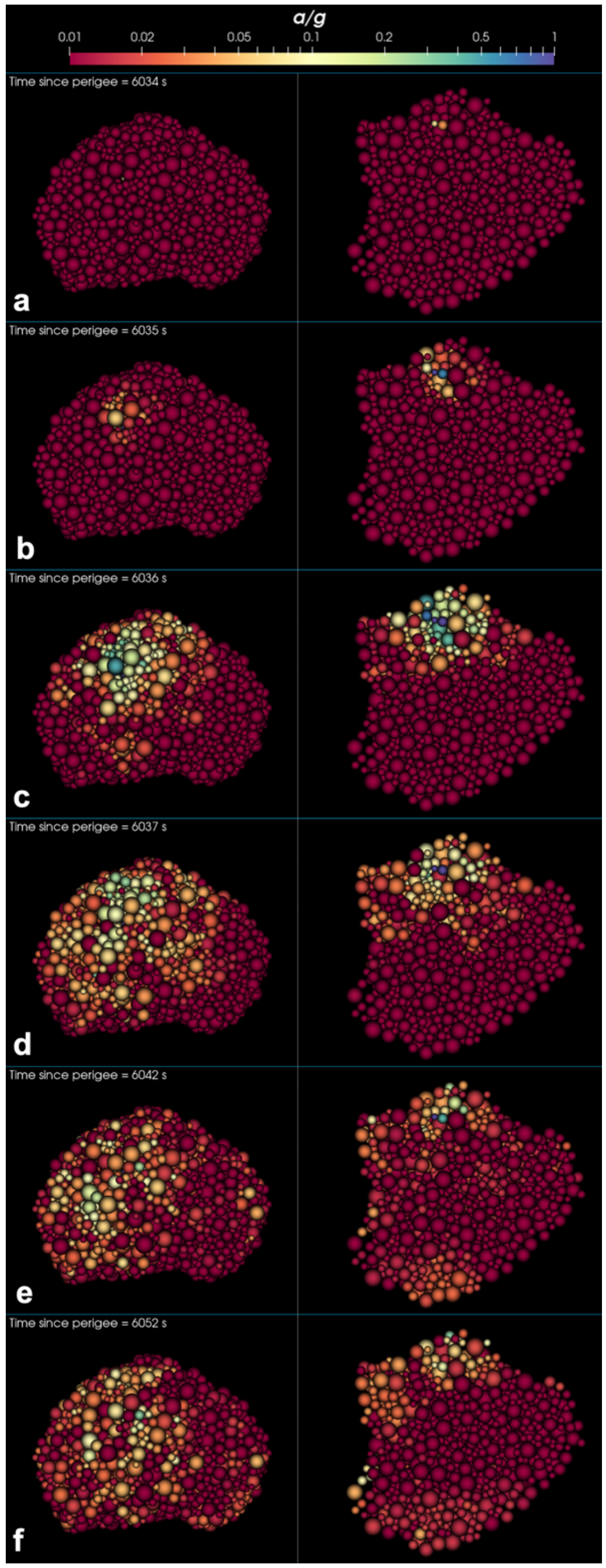

**Figure A3**. Example of a post-perigee internal geotechnical failure that leads to global shaking on Apophis. The left side of each panel is centered with the +y-axis of Apophis pointing out of the page, providing a view of the surface propagation of the seismic wave. The right side of each panel shows a slice along the x-y plane of Apophis, with the +z-axis of Apophis pointing into the page. **a,** 6034 s after perigee, a region of Apophis that is ~40-m below the surface begins to fail as two particles (seen in the right panel) experience sufficiently large shearing forces as two trigger a shaking event that propagates to the surface and through the asteroid. **b, a** + 1 s. **c**, **a** + 2 s. **d, a** + 3 s. **e**, **a** + 8 s. **f, a** + 18 s. See Supplementary Movie M1 for an animation of the entire ~2-minute event.

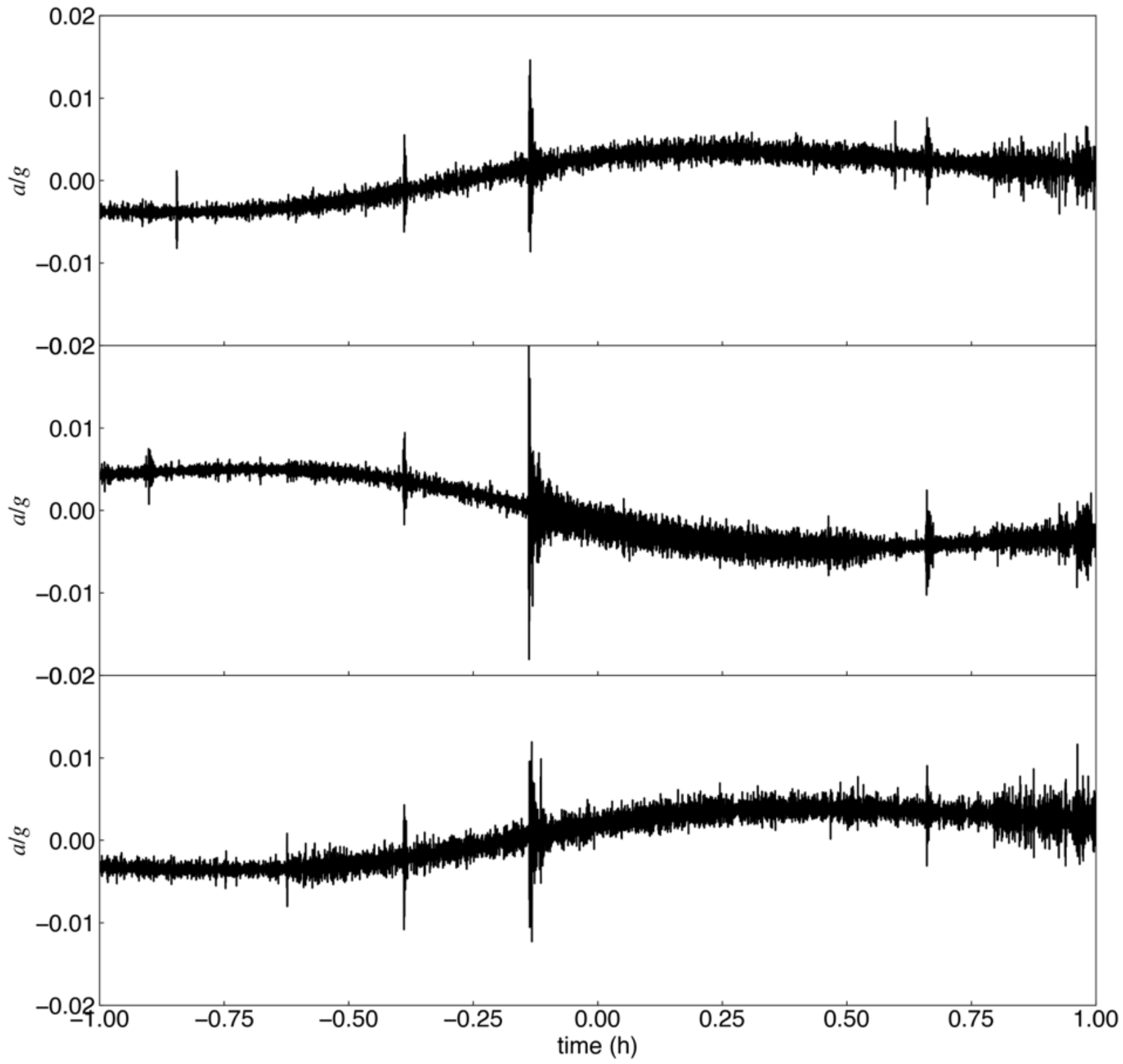

**Figure A4.** The radial acceleration for the surface particles highlighted in Fig. 3d (*top panel*), Fig. 3e (*middle panel*), and Fig. 3f (*bottom panel*) for -1 to 1 h since perigee. The *top and bottom panels* show the axes that are originally in compression (negative values of acceleration), due to tidal stresses, begin to relax as the asteroid moves past Earth. The surface particle along the orthogonal axis (*middle panel*) is initially in tension (positive values). This latter surface particle is aligned with the vector that points from Apophis to Earth in that simulation. As Apophis flies by Earth, the alignment of different body axes with the tidal force flips. Axes that were once in tension become compressed and vice-versa.